\documentclass{aa}

\usepackage{amsmath,amsfonts,amssymb} %for math
\usepackage{bm} %for vectors
\usepackage[table,  dvipsnames]{xcolor} %for author comment colors and that table
\usepackage{hyperref} %to edit citation properties
\usepackage{nicefrac} %for some nice inline fractions
\usepackage{placeins}
\usepackage{microtype}
\usepackage{siunitx}
\usepackage{lipsum}

\usepackage[varg]{txfonts} %aa requirement
\usepackage[T1]{fontenc}

\usepackage{soul}

\hypersetup{
	pdftitle = {Planet formation at the inner disk edge: outbursts, rings, vortices, and turbulence},
	pdfauthor = {A. Ziampras, T. Birnstiel},
	pdfsubject = {},
	colorlinks = {true},
	linkcolor = [rgb]{0,0.35,0.7},
	citecolor = [rgb]{0,0.35,0.7},
	filecolor = [rgb]{0.61,0,0},
	urlcolor = [rgb]{0.61,0,0},
}

\usepackage{graphicx}  %for images
\graphicspath{{./}}

\newcommand{\DP}[2]{\frac{\partial{#1}}{\partial{#2}}}

\newcommand{\Mstar}{M_\star}

\newcommand{\Msun}{\mathrm{M}_\odot}

\newcommand{\Mearth}{\mathrm{M}_\oplus}

\newcommand{\cs}{c_\mathrm{s}}

\newcommand{\vel}{\bm{u}}

\newcommand{\vfrag}{v_\mathrm{frag}}

\newcommand{\Sigmag}{\Sigma_\mathrm{g}}
\newcommand{\Sigmad}{\Sigma_\mathrm{d}}

\newcommand{\velg}{\vel_\mathrm{g}}

\newcommand{\St}{\mathrm{St}}

\newcommand{\amax}{a_\mathrm{max}}

\newcommand{\TMRI}{T_\text{MRI}}
\newcommand{\alphaMRI}{\alpha_\text{MRI}}
\newcommand{\alphaDZ}{\alpha_\text{DZ}}

\newcommand{\Mdotacc}{\dot{M}_\text{acc}}

\newcommand{\pluto}{\texttt{PLUTO}}

\newcommand{\optool}{\texttt{optool}}

\newcommand{\tripod}{\texttt{TriPoD}}

\defcitealias{ziampras-etal-2026}{paper I}

\begin{document}

\title{Planet formation at the inner edge of the dead zone}
\subtitle{II. Outbursts, rings, vortices, and suppression of planetesimal formation}
\titlerunning{Accretion outbursts and vortices}
\author{
	Alexandros~Ziampras\thanks{E-mail: a.ziampras@lmu.de}\inst{\ref{LMU},\ref{MPIA}}
	\and Tilman~Birnstiel\inst{\ref{LMU},\ref{ORIGINS}}
}

\institute{
	Ludwig-Maximilians-Universit{\"a}t M{\"u}nchen, Universit{\"a}ts-Sternwarte, Scheinerstr.~1, 81679 M{\"u}nchen, Germany\label{LMU}
    \and Max Planck Institute for Astronomy, K{\"o}nigstuhl 17, 69117 Heidelberg, Germany\label{MPIA}
	\and Exzellenzcluster ORIGINS, Boltzmannstr.~2, 85748 Garching, Germany\label{ORIGINS}
}

\date{\today}

\abstract
	{~Accretion outbursts have been observed in a variety of young stellar objects, but models of their dynamical evolution have been largely limited to axisymmetric models due to their computational cost.
	}
	{~We investigate the azimuthal stability of accretion outbursts 
	and the formation of planetesimals during these events.
	}
	{~We performed high-resolution 2D, vertically integrated multifluid radiation-hydrodynamical simulations of the inner $\sim\!10$\,au of protoplanetary disks with a dynamically growing dust population, including radiation transport and a realistic dust opacity model.
	}
	{~Accretion outbursts are highly unstable to the Rossby-wave instability, with the burst front quickly diffusing into a large number of small-scale vortices that coalesce over time into a single, compact vortex and inducing azimuthal asymmetries. Vortices act as a source of vigorous turbulent diffusion, strongly suppressing planetesimal formation. 
	}
	{~Our results suggest that azimuthal asymmetries associated with accretion outbursts should be both common and detrimental to planet formation. Nevertheless, planetesimal formation will resume post-burst, as the burst-induced vortices eventually decay and the disk returns to a quiescent state featuring a pressure bump at $\sim\!1$\,au.
	}


\bibpunct{(}{)}{;}{a}{}{,}

\maketitle

\section{Introduction}
\label{sec:introduction}

% FU Ori outbursts and the TI
Many young stellar objects exhibit large-amplitude luminosity variability, including a spectrum of EX Lup to FU Ori-type outbursts with brightness increases of several orders of magnitude \citep{hartmann-kenyon-1996,audard-etal-2014,fischer-etal-2023}. These events are widely interpreted as episodic accretion events, with a leading explanation for such outbursts being a thermal instability \citep[TI, e.g.,][]{lin-etal-1985,kley-lin-1999} operating near the transition between magnetically active and inactive regions of the disk \citep{dullemond-monnier-2010}.

% what is the TI
In this picture, the inner disk is sufficiently hot and ionized to sustain turbulence via the magnetorotational instability \citep[MRI,][]{balbus-hawley-1991,hawley-etal-1995} while the outer disk hosts a colder, weakly ionized dead zone with reduced angular momentum transport \citep{gammie-1996,bai-stone-2013b,lesur-etal-2023b}. The resulting mismatch in accretion efficiency leads to mass accumulation at the dead zone inner edge. As the surface density increases, viscous heating can eventually exceed radiative cooling, triggering a runaway transition to a hot, ionized state, activating the MRI and driving a burst of vigorous accretion onto the central star \citep{armitage-etal-2001,zhu-etal-2009}.

% past 1D studies
The evolution of stellar outbursts via the TI has been extensively studied, with the vast majority of models in the literature focused on the radial evolution of the burst in 1D \citep[e.g.,][]{bell-lin-1994,armitage-etal-2001,zhu-etal-2009,bae-etal-2014,chambers-2024,ziampras-etal-2026} or its (still axisymmetric) radial--vertical structure in 2D \citep[e.g.,][]{wunsch-etal-2006,cecil-flock-2024,cecil-etal-2026}. However, the azimuthal stability of the burst front has been largely unexplored, with \citet{cecil-etal-2026} only briefly mentioning that the Rossby-wave instability \citep[RWI,][]{lovelace-1999} criterion should be satisfied across the burst front during its evolution. Investigating the role of the RWI during an accretion outburst is important for a number of reasons, from characterizing the role of vortices in the burst evolution to evaluating their impact on the formation of planetesimals during an outburst event.

% connection to planetesimal formation
In \citet{ziampras-etal-2026} \citepalias[hereafter][]{ziampras-etal-2026}, we investigated the role of dust evolution in stellar outbursts with the first dedicated models of the TI involving both dynamic dust growth and dust--gas aerodynamic and thermal coupling, but restricted our models to 1D axisymmetric due to their physical complexity. There, we concluded that for reasonable levels of turbulent diffusion in the dead zone (with $\alphaDZ=10^{-4}$), the radial substructure induced by an outburst event resulted in multiple prominent pressure bumps, with the outermost one serving as a long-lived trap that could produce up to $\sim\!1.6\,\Mearth$ of planetesimals over 50\,kyr after the burst. Vortices could impact this picture by either enhancing planetesimal formation by concentrating dust in their centers \citep[e.g.,][]{Barge1995,klahr-henning-1997,birnstiel-etal-2013,lyra-lin-2013}, or by suppressing it via turbulent diffusion \citep[e.g.,][]{kuznetsova-etal-2022}.

% what we do in this paper
In this work, we relax the axisymmetric assumption of \citetalias{ziampras-etal-2026} and perform 2D radiation-hydrodynamic simulations of the TI at the dead zone inner edge with a dynamically growing dust population. Our goals are to evaluate the role of the RWI in influencing both the burst evolution and planetesimal formation, measure the turbulent stress that vortices induce during the burst, and evaluate its impact on the formation of substructure and planetesimals. In doing so, we aim to provide a more complete picture of the TI at the dead zone inner edge and its implications for planet formation, until fully 3D multifluid radiation-hydrodynamic models become computationally feasible.

% ... and now section summaries
We provide a brief summary of the physical and numerical framework of our models in Sect.~\ref{sec:physics-numerics}, and then present our results on the accretion outbursts and planetesimal formation in Sect.~\ref{sec:results-outbursts}. We discuss our results in Sect.~\ref{sec:discussion} and summarize our main conclusions in Sect.~\ref{sec:summary}.

\section{Physics and numerics}
\label{sec:physics-numerics}

Our physical framework is identical to that of \citet{ziampras-etal-2026} \citepalias[hereafter][]{ziampras-etal-2026}, and we refer the reader to that paper for details. In short, we solve the 2D, vertically integrated Navier--Stokes equations of hydrodynamics and radiative transfer in a cylindrical coordinate system $\{R,\phi, z=0\}$ for gas with surface density $\Sigmag$ and temperature $T$ and dust with surface density $\Sigmad$ at distance $R$ from a star with mass $\Mstar=1\,\Msun$. We include heating due to compression, turbulent dissipation (modeled as viscosity), and stellar irradiation, cooling via thermal emission through the disk surfaces and in-plane radiative diffusion via flux-limited diffusion \citep[FLD,][]{levermore-pomraning-1981}, a temperature-dependent $\alpha$-viscosity prescription \citep{shakura-sunyaev-1973} that mimics the transition from the active to the dead zone \citep{cecil-flock-2024}, and a state-of-the-art treatment of the dust size distribution including two-way gas--dust drag, dust diffusion, dust coagulation and fragmentation \citep[see][]{pfeil-etal-2024}, and a self-consistent, temperature-dependent opacity model for an evolving dust size distribution.

\subsection{Numerics}
\label{sec:numerics}

We use the \pluto{} \texttt{v.4.4-patch3} code \citep{mignone-etal-2007} to solve the Navier--Stokes equations, including the FARGO algorithm \citep{masset-2000,mignone-etal-2012} to dramatically speed up azimuthal advection. Radiative transfer with FLD is implemented with the implicit solver described in \citet{ziampras-etal-2020a}. Dust is modeled as a pressureless fluid following \citet{ziampras-etal-2025a}, with dust diffusion as in \citet{weber-etal-2019} and dust evolution via coagulation--fragmentation implemented using \tripod{} \citep{pfeil-etal-2024}. The dust opacity model was calculated with \optool{} \citep{dominik-etal-2021} and then tabulated and interpolated on the fly using \texttt{growpacity}\footnote{\url{https://github.com/alexziab/growpacity}} \citep{ziampras-birnstiel-2026}.

Our computational domain extends radially between $R\in[0.05,10]$\,au and covers the full $2\pi$ in azimuth. The radial grid utilizes two logarithmically spaced segments with 1024 cells between 0.05--2\,au and 128 cells between 2--10\,au, while the azimuthal grid consists of 2048 uniformly spaced cells. With a nominal pressure scale height of order $\sim0.025\,R_\text{au}^{9/7}$, this resolution yields approximately 7 cells per scale height at 1\,au during the quiescent phase, enhanced to 21 cells per scale height during the burst phase. Identical to \citetalias{ziampras-etal-2026}, wave-damping zones are implemented near radial boundaries.

% \begin{table*}[b]
% 	\begin{center}
% 	\caption{List of models presented in this paper. All models share $\alphaMRI=0.1$.}
% 	\label{tab:models}
% 	\begin{tabular}{c|c|c|c|l}
% 		\hline
% 		$\alphaDZ$ & $\vfrag$ [m/s] & $N_R\times N_\phi$ & dust backreaction? & motivation \\
% 		\hline
% 		\hline
% 		$10^{-4}$ & 1 & $1152\times2048$ & yes & our fiducial model with a realistic $\alphaDZ$ \\
% 		$10^{-3}$ & 1 & $1152\times2048$ & yes & vortex growth in a more turbulent dead zone \\
% 		$10^{-4}$ & 10 & $1152\times2048$ & yes & testing the effect of very large grains ($\gtrsim 1\,$cm)\\
% 		$10^{-4}$ & 1 & $1152\times2048$ & no & effect of dust backreaction on vortex survival \\
% 		$10^{-4}$ & 1 & $4608\times8192$ & yes & development of the RWI at very high resolution \\
% 		\hline
% 	\end{tabular}
% 	\end{center}
% \end{table*}

Our models are initialized with the same gas and dust surface density profiles as in the 1D models of \citetalias{ziampras-etal-2026}, which are based on the semi-analytical calculations of the pre-burst disk structure therein. This ensures realistic initial conditions, and enables direct comparisons between our 2D models presented here and the 1D models in \citetalias{ziampras-etal-2026}. The 1D radial profiles are then broadcast in $\phi$ and the gas velocity field is perturbed with random noise with an amplitude of 1\% of the local sound speed to break axisymmetry. We evolve the models for a full burst cycle, which depending on model parameters lasts between 70--100\,yr. We denote the time of the burst onset as $t_\text{burst}=0$.

For our fiducial model we use $\alphaMRI=0.1$, $\alphaDZ=10^{-4}$, and a fragmentation velocity $\vfrag=1\,$m/s. We then explore the effect of a more turbulent dead zone with $\alphaDZ=10^{-3}$ and a higher fragmentation velocity of $\vfrag=10\,$m/s. To test the robustness of our results regarding the growth of the RWI and the resulting vortices, we add three more models:
\begin{itemize}
	\item one without dust aerodynamic backreaction, which has been shown to dampen vortex survival in vertically integrated models \citep{raettig-etal-2015,lovascio-etal-2022,ziampras-etal-2025a},
	\item a simpler model where in-plane cooling via FLD is omitted (presented in Appendix~\ref{apdx:noFLD}) to test for the effect of radiative diffusion on the impact and lifetime of vortices, and
	\item a high-resolution model to check for numerical convergence as well as to zoom in on the early development of the RWI.
\end{itemize}
We therefore run a total of six models, where the first five models use the grid parameters described above, and our final model is run at quadruple resolution with $N_R\times N_\phi = (4096+512)\times8192 = 4608\times8192$, for an effective resolution of 28 cells per scale height at 1\,au during the quiescent phase and 84 cells per scale height during the burst phase. %A summary of our models is given in Table~\ref{tab:models}.

\section{Results: accretion outbursts}
\label{sec:results-outbursts}

In this section we present the results of our hydrodynamical simulations regarding the evolution of the accretion outburst, the growth of vortices, and the associated turbulent stress. We will mainly focus on our fiducial model with $\alphaDZ=10^{-4}$ and $\vfrag=1\,$m/s, and refer to other models when relevant differences arise. For this reason, not all quantities are shown for every model.

\subsection{Burst evolution} \label{sub:burst-evolution}

\begin{figure*}
	\includegraphics[width=\textwidth]{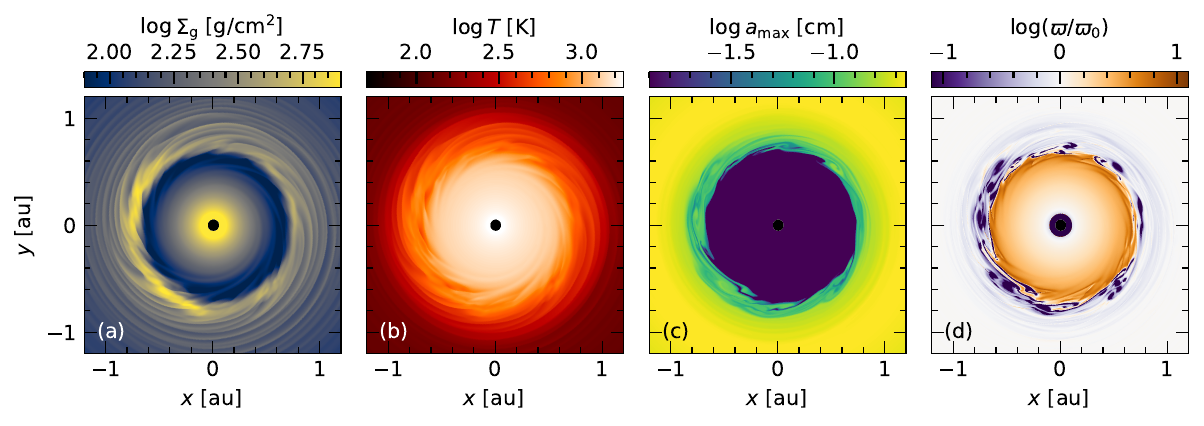}
	\caption{Snapshot of our fiducial model ($\alphaDZ=10^{-4}$, $\vfrag=1\,$m/s) at $t=20$\,yr, during the burst phase. Panels show the gas surface density $\Sigmag$ (\emph{a}), temperature $T$ (\emph{b}), maximum grain size $\amax$ (\emph{c}), and normalized vortensity $\varpi/\varpi_0$ (\emph{d}), with minima in the latter denoting the presence of vortices. The disk structure is highly nonaxisymmetric, with a dense crescent in $\Sigmag$.
	}
	\label{fig:snapshot-2D}
\end{figure*}

Once the burst ignites at the dead zone inner edge, the burst front begins propagating outward as the MRI-active region expands radially. Figure~\ref{fig:snapshot-2D} shows a snapshot of our fiducial model at $t=20$\,yr, where the burst front has reached its outermost radius at $R\approx0.75$\,au, highlighting the nonaxisymmetric nature of the flow with a dense crescent in the gas surface density $\Sigmag$ (panel \emph{a}) and numerous spirals in $\Sigmag$ and the temperature $T$ (panel \emph{b}). This behavior owes to the development of the RWI along the burst front during its outward propagation, which quickly breaks up the front into a large number of small-scale vortices, evident in the vortensity $\varpi=\hat{z}\cdot(\nabla\times\velg)/\Sigmag$ (panel \emph{d}, normalized to its unperturbed profile $\varpi_0$). These vortices in turn induce significant radial and azimuthal mixing, smoothing and broadening the burst front radially. This can be seen in the maximum grain size $\amax$ (panel \emph{c}), which, in stark contrast to the sharp radial transition from $\sim\!10\,\mu$m to $\sim\!3$\,mm in the 1D models of \citetalias{ziampras-etal-2026}, shows a much more gradual transition in addition to swirling patterns due to the vortices.

\begin{figure}
	\includegraphics[width=\columnwidth]{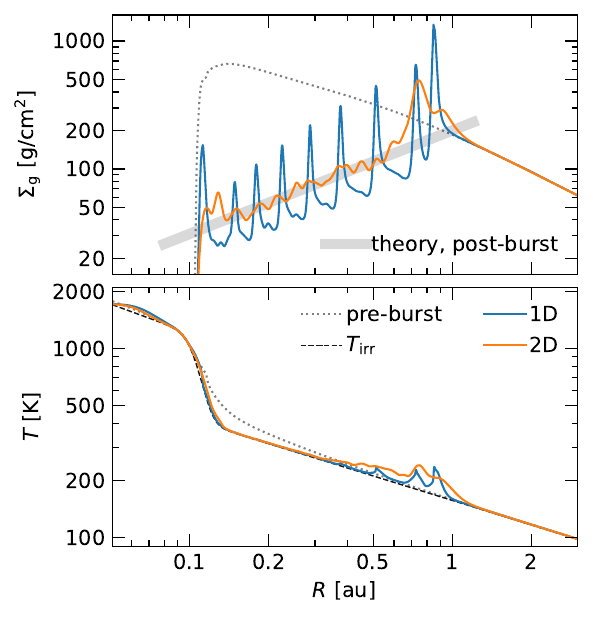}
	\caption{Comparison between our fiducial model (orange lines) and an equivalent 1D model \citepalias[blue, see][]{ziampras-etal-2026} at the end of the burst phase. The series of spikes in 1D gives way to smeared out radial profiles in the burst region in 2D. The analytical prediction for the shape of the post-burst region from \citetalias{ziampras-etal-2026} is shown with a thick gray line, and is in excellent agreement with the 2D model results.
	}
	\label{fig:1D-vs-2D_a4}
\end{figure}

\begin{figure}
	\includegraphics[width=\columnwidth]{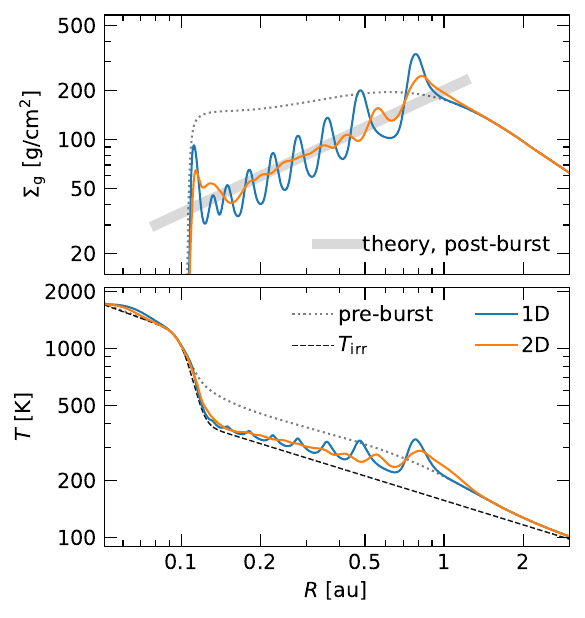}
	\caption{Similar to Fig.~\ref{fig:1D-vs-2D_a4} for our more turbulent model with $\alphaDZ=10^{-3}$. The same overall behavior and good agreement with the analytical predictions is observed, albeit with less pronounced features due to the higher turbulent diffusion.}
	\label{fig:1D-vs-2D_a3}
\end{figure}

A more quantitative comparison between our fiducial 2D model and an equivalent 1D model akin to \citetalias{ziampras-etal-2026} is shown in Fig.~\ref{fig:1D-vs-2D_a4}, at a time corresponding to the end of the burst phase with $t_\text{2D}=98$\,yr and $t_\text{1D}=126$\,yr. It becomes immediately clear that the serrated structure of the region subject to the outburst due to reflares in the 1D model is completely absent and smoothed out in the 2D model, which instead shows a single, broad density bump with a maximum at $R\approx0.75$\,au and a smooth, albeit still inverted, density profile within (top panel). Similarly, the temperature profile shows a smoother, wider hot lump colocated with the density bump in 2D instead of several sharp peaks in 1D (bottom panel).

A very similar behavior is observed in the more turbulent model with $\alphaDZ=10^{-3}$, shown in Fig.~\ref{fig:1D-vs-2D_a3} (with $t_\text{2D}=70\,$yr and $t_\text{1D}=76$\,yr), albeit with overall less pronounced features in both 1D and 2D due to the higher turbulent diffusion. Reassuringly, the analytical predictions for the shape of the post-burst region from \citetalias{ziampras-etal-2026} (thick gray lines) are in excellent agreement with the 2D model results in both cases, confirming the validity of the analytical model and its underlying assumptions.

\begin{figure}
	\includegraphics[width=\columnwidth]{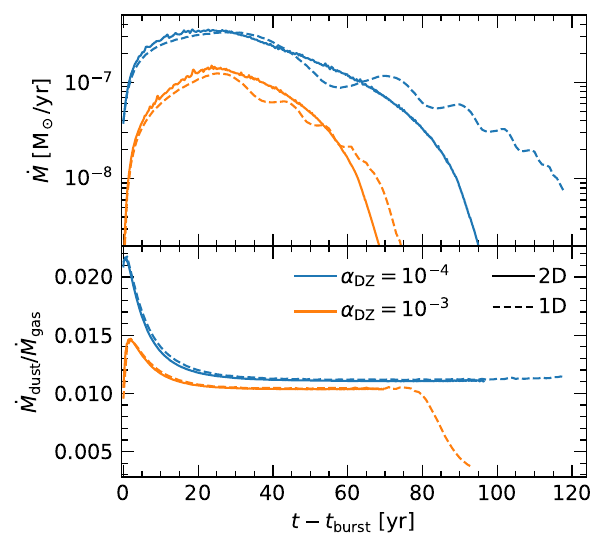}
	\caption{Top: accretion rate $\Mdotacc$ onto the star as a function of time for our 2D models (solid lines) with $\alphaDZ=10^{-4}$ (blue) and $\alphaDZ=10^{-3}$ (orange), compared to their 1D counterparts (dashed lines). Both 1D and 2D models reach similar peak accretion rates, but 2D models exhibit a single accretion spike and a shorter burst duration. Bottom: the dust-to-gas ratio of accreted material through the inner boundary, increasing during the early stages of the burst for both 1D and 2D models.}
	\label{fig:mdot-1D-vs-2D}
\end{figure}

As in \citetalias{ziampras-etal-2026}, the burst is accompanied by a significant increase in the accretion rate $\Mdotacc$ onto the star, computed as
\begin{equation}
	\label{eq:mdot}
	\Mdotacc = -2\pi R \left[\Sigmag {u_R}_\text{g} + \sum\limits_{i}{\Sigma_{\text{d},i}} {u_R}_{\text{d},i}\right]\Big|_{R=R_\text{in}}
\end{equation}
and shown in the top panel of Fig.~\ref{fig:mdot-1D-vs-2D} for both pairs of models. Here, we observe that while both 1D and 2D models reach very similar peak accretion rates, the 2D models show a single, broader accretion spike, rather than multiple undulations seen in 1D, and the burst duration is noticeably shorter in 2D. This is a direct consequence of the aggressive mixing and smoothing of the burst front by the RWI, washing out the reflares and leading to steadier accretion. It is also worth noting that since the burst travels less far out in the 2D model (0.75\,au instead of 0.9\,au in 1D), it is rather unsurprising that the burst duration is shorter in 2D, as the burst region spans a smaller radial extent and thus contains less mass to be accreted. In the bottom panel of the same figure we also show the dust-to-gas ratio of accreted material, which increases by a factor of 1.5--2 depending on $\alphaDZ$ during the early phase of the burst, possibly related to observational evidence for enhanced metallicity during outburst events \citep[e.g.,][]{guenther-etal-2018}.

All in all, the burst evolution is qualitatively similar between 1D and 2D, albeit with a much smoother radial structure due to the formation of vortices and the resulting turbulent mixing. In the next section we will investigate the growth of the RWI in more detail.

\subsection{Vortex formation} \label{sub:vortex-formation}

\begin{figure}
	\includegraphics[width=\columnwidth]{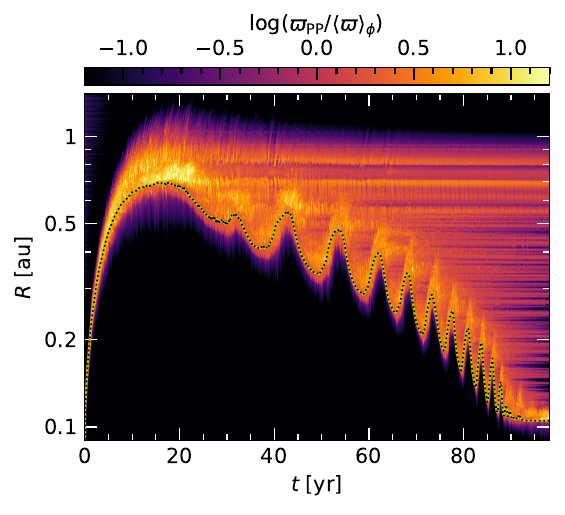}
	\caption{Radius--time heatmap of the peak-to-peak vortensity contrast $\varpi_\text{pp}/\langle\varpi\rangle_\phi$ for our high-resolution model. Bright features indicate nonaxisymmetric vortex-like features, whilst black regions represent an axisymmetric state. A dotted black line traces the location of the burst front by proxy of the radius where $T=\TMRI=900\,$K. More than ten reflares are resolved in this figure, each giving rise to more vortices, with those in the outer disk outliving both those in the inner disk and the burst itself.}
	\label{fig:a4-vortensity-Rt}
\end{figure}

We now move on to quantify the radial and temporal extent to which RWI-driven vortices develop, act, and survive in our models. Here, for illustrative purposes, we will use the high-resolution variant of our fiducial model ($\alphaDZ=10^{-4}$, $\vfrag=1$\,m/s, $4\times$ the resolution along both axes) as it better highlights the development of the RWI while still showing the same overall behavior as the fiducial model. A comparison between the two can be found in Appendix~\ref{apdx:highres}.

Figure~\ref{fig:a4-vortensity-Rt} shows a radius--time diagram of the peak-to-peak vortensity contrast $\varpi_\text{pp} = \max_\phi(\varpi) - \min_\phi(\varpi)$ at a given radius, normalized to the azimuthal median $\langle\varpi\rangle_\phi$, for our high-resolution model. Here, bright features indicate the presence of nonaxisymmetric vortex-like structures, with the brightness related to their amplitude. We find that the flow becomes unstable to the RWI as soon as the burst front starts propagating outward at $R\sim0.1$\,au. Interestingly, even though the radial disk profile is quite smooth in 2D, a number of clearly defined reflares can be seen in this figure, with each reflare giving rise to a new generation of vortices and partially refreshing existing ones. Finally, we find that vortices in the outer disk ($R\gtrsim0.5$\,au) are more long-lived than those in the inner disk, surviving for the entire burst phase, while inner vortices are more short-lived and transient, likely due to the shorter orbital and viscous timescales in the inner disk.

\begin{figure}
	\includegraphics[width=\columnwidth]{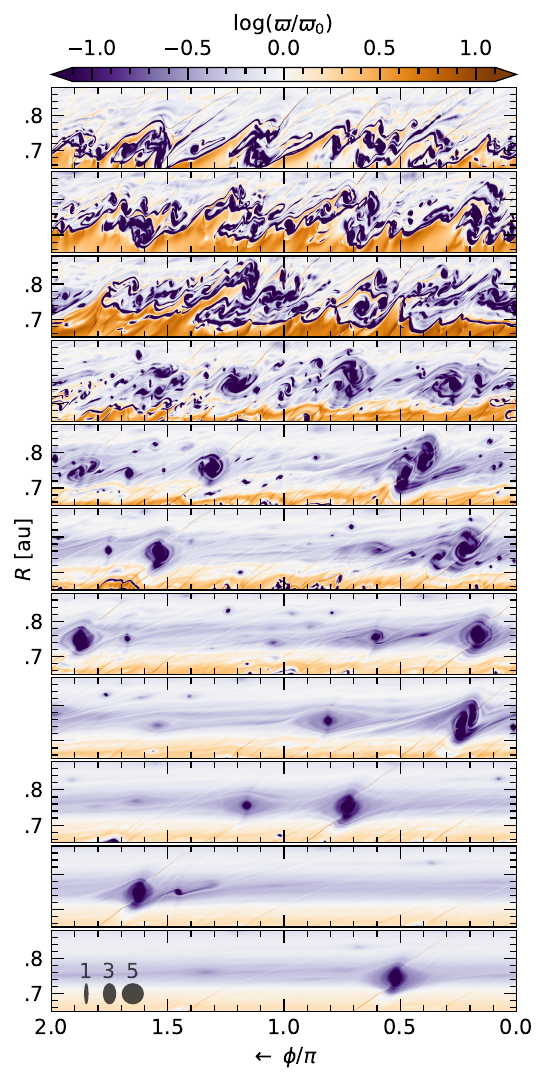}
	\caption{Heatmaps of the perturbed vortensity in our high-resolution model over several snapshots, capturing the formation of vortices as the burst front reaches $R\sim0.75$\,au (panels \emph{a}--\emph{c}) and their subsequent merging (\emph{d}--\emph{j}) until a single vortex has survived (panel \emph{k}). Ellipses on panel \emph{k} illustrate elliptical features (e.g., vortices) with a given aspect ratio $\chi\equiv R\Delta\phi/\Delta R$.}
	\label{fig:vortensity-2D-vertical}
\end{figure}

To highlight the formation of vortices in more detail we plot the vortensity field for several snapshots over the course of the burst evolution near $R\sim0.75$\,au, where the burst front stops and recedes, in Fig.~\ref{fig:vortensity-2D-vertical}. From top to bottom, we see the burst front arrive at $R\sim0.75$\,au (panels \emph{a}--\emph{c}), and the subsequent formation of a large number of small-scale vortices over a radial extent of $\sim0.2$\,au. As the burst front recedes (panel \emph{d}), the vortices begin to coalesce, with multiple merger events observed in panels \emph{e}--\emph{j}, eventually leading to a single surviving vortex before the end of the burst phase (panel \emph{k}). These vortices are rather compact, with the resulting vortex in panel \emph{k} having an aspect ratio $\chi=R\Delta\phi/\Delta R\approx5$ (measured by approximating the vortex with a Gaussian and evaluating its full width at half maximum in both directions). This likely relates to the very strong vortensity contrast at the burst front, which results in vortices with a very low Rossby number and thus a more compact structure \citep[see, e.g.,][]{hammer-lin-2023}.

This process of vortex formation and coalescence is both observed in all our models, suggesting that the RWI is a robust outcome of accretion outbursts, and reoccurs during every reflare in the burst phase, which is a direct consequence of the fact that each reflare shapes a new, sharp vortensity spike along the burst front. These observations, combined with the resulting smoother radial structure of the post-burst disk region, suggest that these vortices can drive turbulent mixing and diffusion with a diffusion coefficient much higher than $\alpha\sim10^{-3}$ (the case of our most viscously diffusive model). We quantify this vortex-driven turbulent stress in the next section.

\subsection{Vortex-driven turbulent stress}
\label{sub:vortex-stress}

At this point, it is clear that the RWI and the resulting vortices play a crucial role in shaping the burst front and its evolution, providing significant turbulent mixing and diffusion via both the growth of the RWI and their launching of spiral density waves \citep[akin to][]{goodman-rafikov-2001}. To quantify the strength of this vortex-driven turbulent stress, we compute an effective ``turbulent viscosity'' parameter $\alpha_\text{turb}$ via the Reynolds stress following \citet{balbus-papaloizou-1999}:
\begin{equation}
	\label{eq:alpha-vort}
	\!\!\alpha_\text{turb}(R,\phi,t) = \left|\DP{\log\bar{\Omega}_\phi}{\log R}\right|^{-1}\!\!\frac{\langle\Sigmag \delta u_R \delta u_\phi\rangle}{\langle P \rangle}, \!\!\quad\delta u_x \!= \!u_x - \frac{\oint{u_x\!\Sigmag \mathrm{d}\phi}}{\oint{\Sigmag \mathrm{d}\phi}}.
\end{equation}
In the above equation, $\bar{\Omega}_\phi$ is the azimuthally averaged angular velocity, $\langle\cdot\rangle$ denotes smoothing using a rolling average\footnote{We use a Savitzky--Golay filter (\texttt{scipy.signal.savgol\_filter} in \texttt{python}), with a polynomial order of 1.} with a window size of 2 scale heights in both $R$ and $\phi$, and $\delta u_x$ is the velocity fluctuation in the $x$ direction after subtracting the density-weighted azimuthal average at a given radius. This definition of $\delta u_x$ has been shown to yield a more accurate estimate of the residual velocities in the presence of a nonaxisymmetric background flow \citep{cordwell-etal-2025}.

\begin{figure*}
	\includegraphics[width=\textwidth]{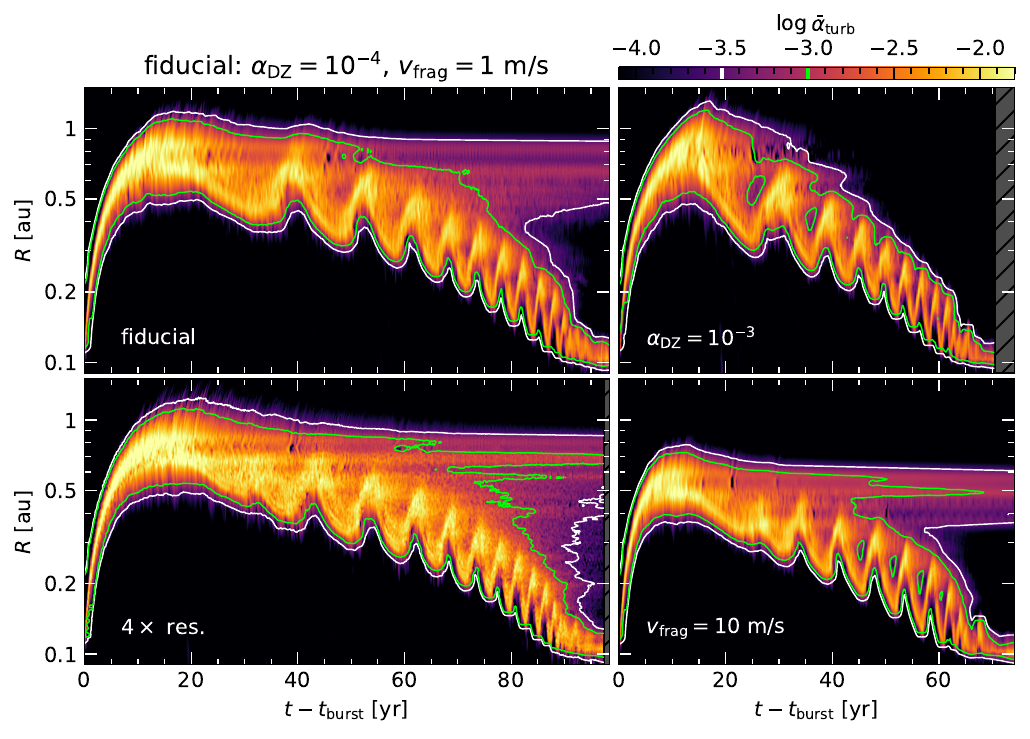}
	\caption{Radius--time heatmaps of the azimuthally averaged turbulent viscosity $\bar{\alpha}_\text{turb}(R,t)$ for (top to bottom, left to right) our fiducial model with $\alphaDZ=10^{-4}$ and $\vfrag=1\,$m/s, a more diffusive model with $\alphaDZ=10^{-3}$, our high resolution model, and one with a higher fragmentation velocity of $\vfrag=10\,$m/s. As in Fig.~\ref{fig:a4-vortensity-Rt}, bright colors denote RWI-driven turbulence. Green and white lines trace contours where $\bar{\alpha}_\text{turb}=10^{-3}$ and $3\times10^{-4}$, respectively, as qualitative indicators of substantial turbulence.}
	\label{fig:alpha-Rt-gallery}
\end{figure*}

We then show the resulting azimuthally averaged $\bar{\alpha}_\text{turb}(R,t)$ for our fiducial, high-viscosity, high-resolution, and high-fragmentation-velocity models in Fig.~\ref{fig:alpha-Rt-gallery}. Similar to Fig.~\ref{fig:a4-vortensity-Rt}, bright features indicate the presence of turbulence, here primarily driven by the RWI along the burst front. Notably, this illustration also quantifies the turbulent stress driven by vortices as they decay, long after the burst front has receded from a given radius: vortices dissipate nearly immediately due to viscous diffusion for $\alphaDZ=10^{-3}$, while they survive for at least the entire burst duration for $\alphaDZ=10^{-4}$, albeit only down to $R\approx0.5$\,au. As vortices are both better resolved and less exposed to numerical diffusion in our high-resolution model, they feature both a higher overall amplitude and a longer lifetime, but still decay on burst timescales in the inner disk, suggesting that our models are reasonably well converged in terms of vortex behavior.

An additional feature of the high-resolution model is that the first reflare (starting at $t\approx 35$\,yr in our fiducial model) is delayed until $t\approx 40$\,yr instead. Similar behavior was observed in our model without FLD (see Fig.~\ref{fig:alpha-Rt-noFLD} and Appendix~\ref{apdx:noFLD}), with the first reflare starting at $t\approx 45$\,yr there. This points to the possibility that a high enough vortex-driven stress can interfere with the burst evolution itself, delaying the onset of reflares to an extent. In our models this was not significant enough to change the overall burst cycle, but we cannot guarantee that this would not be the case at much, much higher resolution. At the same time, however, such an incredibly high resolution model is both currently out of reach and likely not very well physically motivated, as we are already resolving sub-scale-height features that could be affected by 3D processes \citep[in a similar vein to][]{mcnally-etal-2019a}.

\begin{figure}
	\includegraphics[width=\columnwidth]{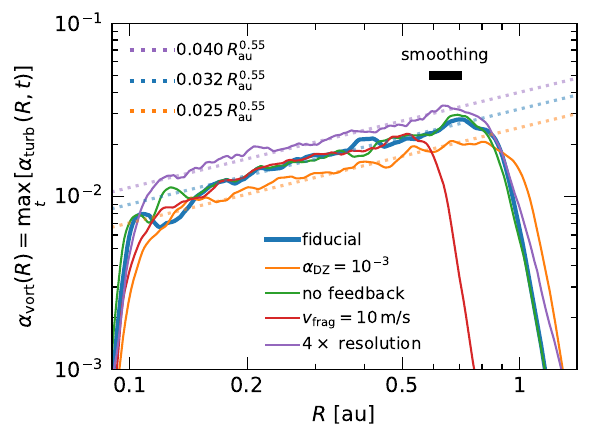}
	\caption{The maximum value of the azimuthally averaged Reynolds stress across the burst phase, as a tracer of the overall strength of vortex-driven turbulent stress. Dotted lines shows power-law fits to the data, with $\alpha_\text{vort}\propto R^{0.55}$. The data is further smoothed with a rolling average (its window denoted by a black bar) for clarity.}
	\label{fig:alpha-time}
\end{figure}

After computing this space- and time-dependent $\alpha_\text{turb}$, we then estimate the overall strength of the vortex-driven turbulent stress by taking its azimuthal average for each snapshot (sampled with a cadence of 0.1\,yr) and then taking the maximum value across the time dimension during the burst phase. In doing so, we implicitly assume that RWI-related stress is both always active and dominant during the burst phase, which is a reasonable assumption  given the behavior seen in Fig.~\ref{fig:alpha-Rt-gallery}. We refer to this reduction of $\alpha_\text{turb}(R,\phi,t)$ as $\alpha_\text{vort}(R)$:
\begin{equation}
	\label{eq:alpha-vort}
	\alpha_\text{vort}(R) = \max_t[ \bar{\alpha}_\text{turb}(R,t)],
\end{equation}
and plot its resulting radial profile in Fig.~\ref{fig:alpha-time}. We find that $\alpha_\text{vort}$ reaches values of order $\gtrsim10^{-2}$ over the burst region even for our turbulent model with $\alphaDZ=10^{-3}$ (albeit with a slightly lower amplitude compared to the fiducial case), and is generally higher in the outer disk, where vortices are also more long-lived. Regarding our other models with $\alphaDZ=10^{-4}$, we find that:

\begin{itemize}
	\item The model without dust aerodynamic backreaction shows practically identical peak values of $\alpha_\text{vort}$, implying that the Rossby-wave instability and the resulting vortices are hardly affected by dust backreaction. This most likely happens due to the very low Stokes numbers of the dust grains ($\St_\text{max}\approx3\times10^{-3}$) and the modest increase in dust-to-gas ratios ($\Sigmad/\Sigmag\lesssim0.05$) in the burst region, which are insufficient to drive the strong dust feedback that can dampen vortices in other contexts \citep{raettig-etal-2015,lovascio-etal-2022,ziampras-etal-2025a}.
	\item The model with $\vfrag=10\,$m/s shows the same overall behavior as the fiducial model, but $\alpha_\text{vort}$ is truncated less far out. This happens since the burst front only propagates out to $\approx\!0.55$\,au, due to the dust growing to much larger sizes ($\sim\!10$\,cm as opposed to $\sim\!1$\,mm), which in turn leads to a much lower opacity and thus more efficient cooling, shutting off the burst at smaller radii \citepalias[in line with our findings in][]{ziampras-etal-2026}.
	\item The high-resolution model shows a higher overall amplitude by a factor of $\sim25\%$ compared to the fiducial model, due to the presence of small-scale vortices that are not fully resolved in the fiducial model (see also Fig.~\ref{fig:vortensity-2D} in Appendix~\ref{apdx:highres}), but with the same overall shape and radial dependence of $\alpha_\text{vort}$.
\end{itemize}

Interestingly, the radial profile of $\alpha_\text{vort}$ is very well fit by a power law with $\alpha_\text{vort}\propto R^{0.55}$, truncated at the inner and outer edges of the burst region. While this result seems promising at face value, as it suggests that the vortex-driven turbulent stress could be reasonably well approximated by a simple power law in future 1D models, we stress that it is not clear whether this behavior is a general feature of the RWI or a coincidence in our particular model. Further investigation is needed to confirm this result and its potential implications for modeling accretion outbursts in 1D.

Overall, regarding the evolution of the burst cycle itself, we find that the inclusion of the azimuthal dimension enables the vigorous growth of the Rossby-wave instability, dramatically changing the structure of the burst front from a series of serrated peaks in 1D to a broad density bump in 2D enclosing a smooth profile. Variations in the accretion rate are also smoothed out as a direct consequence of the above.
At the same time, the overall post-burst structure and the duration of the burst cycle are not significantly affected, and the analytical predictions from \citetalias{ziampras-etal-2026} for the shape of the post-burst region are in excellent agreement with our 2D results. This suggests that the expectation of an inverted density profile in the post-burst region, and therefore the presence of a pressure maximum at $R\sim0.8$\,au within the dead zone, is robust against vortex activity, implying that dust trapping and planet formation in the post-burst region could still be operating. We will investigate this possibility in more detail in the next section.

\subsection{Dusty features and planetesimal formation}
\label{sub:dusty-features}

In \citetalias{ziampras-etal-2026} we showed that the multiple pressure bumps in the post-burst region can act as efficient dust traps, eventually leading to the formation of planetesimals via the streaming instability. In particular, the outermost pressure maximum at $R\sim0.9$\,au for $\alphaDZ=10^{-4}$ was the most relevant dust trap, as it lasted significantly longer than the rest and had direct access to the dust reservoir drifting in from the outer disk. As a result, we had found that over $\sim\!50$\,kyr of post-burst evolution and using the streaming instability criterion of \citet{Lim2025}, a total of $\sim1.6\,\Mearth$ worth of planetesimals had accumulated within the region between 0.1--1.2\,au.

Judging by the shape of the post-burst density profile in Fig.~\ref{fig:1D-vs-2D_a4}, we expect that the outermost pressure maximum at $R\sim0.75$\,au in our fiducial 2D model should still survive for tens of kyr and act as an efficient dust trap for drifting dust grains from the outer disk. This expectation is backed by the fact that the vortices formed during the burst phase will dissipate on timescales much shorter than even 1\,kyr for our choice of $\alphaDZ$ \citep{rometsch-etal-2021}, something we can already see in Fig.~\ref{fig:a4-vortensity-Rt}, and thus the dead zone should exhibit very low levels of turbulent diffusion (here $\alphaDZ=10^{-4}$) during the post-burst quiescent phase, effectively acting as an axisymmetric model. As a result, we expect that our findings in \citetalias{ziampras-etal-2026} regarding the formation of planetesimals via the streaming instability in the post-burst region should still hold in 2D.

However, this is not the case for planetesimals that could have formed during the burst phase itself. Here, two competing mechanisms are at play: azimuthal dust trapping within vortices, which can significantly enhance the dust-to-gas ratio and therefore trigger the streaming instability, and strong turbulent diffusion driven by the RWI that can prevent dust from concentrating, thus suppressing planetesimal formation. To investigate this question, we compare the dust-to-gas ratio $Z=\Sigmad/\Sigmag$ in our models with the critical dust-to-gas ratio for triggering the streaming instability $Z_\text{crit}$ from \citet{Lim2025}
\begin{equation} \label{eq:Zcrit}
	\log Z_\text{crit} = 0.1(\log\St_\text{max})^2 +0.07\log\St_\text{max} - 2.36,
\end{equation}
\begin{figure}
	\includegraphics[width=\columnwidth]{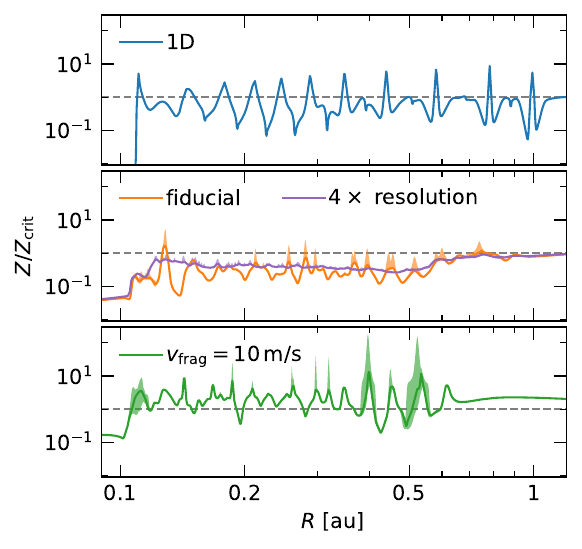}
	\caption{The dust-to-gas ratio $Z$ compared to the critical value needed to trigger the streaming instability, $Z_\text{crit}$, for different models. Regions where $Z/Z_\text{crit}>1$ (above the dashed gray line) are susceptible to planetesimal formation. The latter is heavily suppressed in our high-resolution 2D model, but could operate if dust can grow to sufficiently large grains.}
	\label{fig:Z}
\end{figure}
where $\St_\text{max} = \frac{\pi}{2}\nicefrac{\amax \tilde{\rho}}{\Sigmag}$ is the Stokes number of the largest grains, with $\tilde{\rho}=2.08$\,g/cm$^3$ their bulk density. This comparison is shown in Fig.~\ref{fig:Z} at the end of the burst phase for our fiducial (2D) model, its 1D and high-resolution counterparts, and the model with $\vfrag=10\,$m/s\footnote{The model with $\alphaDZ=10^{-3}$ is omitted as we find that $Z\lesssim0.1\,Z_\text{crit}$, as in \citetalias{ziampras-etal-2026}.}. From this figure, we draw the following conclusions:
\begin{itemize}
	\item The streaming instability criterion by \citet{Lim2025}, while satisfied at the peaks of the (axisymmetric) pressure bumps in our 1D model, is now met only partially in our 2D fiducial model, and only within vortices. This is a direct consequence of the strong turbulent diffusion driven by the RWI, and suggests that planetesimal formation is likely quenched during the burst phase itself.
	\item The high-resolution model supports the above conclusion, as the ratio $Z/Z_\text{crit}$ is even lower, almost never satisfying the streaming instability criterion even within vortices, except for a tiny peak at the outermost pressure bump at $R\approx0.75$\,au. This happens due to the combination of more, smaller vortices and a higher overall amplitude of $\alpha_\text{vort}$ (see Fig.~\ref{fig:alpha-time}), which together lead to both less efficient azimuthal trapping and stronger turbulent diffusion.
	\item The model with $\vfrag=10\,$m/s shows a much higher ratio of $Z/Z_\text{crit}$, sometimes by a factor of 100, with the streaming instability criterion satisfied over a much wider spatial range. This is expected, as dust grains now reach $\St_\text{max}\sim0.1$, making both radial and azimuthal trapping much more efficient. The SI criterion is in fact satisfied even in the outer disk, suggesting a continuous stream of planetesimals forming unrelated to the burst itself, in line with other works that have investigated this scenario \citep[e.g.,][]{drazkowska-etal-2016}.
\end{itemize}
We can further quantify this planetesimal formation ``efficiency'' by computing the mass of dust that satisfies the streaming instability criterion at the end of the burst,
\begin{equation} \label{eq:mdot-SI}
	M_\text{plan} = \int_{\phi=0}^{2\pi} \int_{R=0.1\,\text{au}}^{1.2\,\text{au}} \left[\max\left(Z-Z_\text{crit}, 0\right)\Sigmag\right] R\,\mathrm{d}R\,\mathrm{d}\phi,
\end{equation}
assuming that all dust that satisfies $Z\geq Z_\text{crit}$ is converted to planetesimals and that the streaming instability is a continuous process (i.e., it stops when $Z\leq Z_\text{crit}$ as planetesimals are drawn from the dust mass). To estimate an upper limit to planetesimal formation instead, we also compute the total mass in planetesimals as
\begin{equation} \label{eq:mdot-SI-upper}
	M_\text{plan}^\text{upper} = \int_{\phi=0}^{2\pi} \int_{R=0.1\,\text{au}}^{1.2\,\text{au}} \Sigma_\text{d} \Big|_{Z\geq Z_\text{crit}} R\,\mathrm{d}R\,\mathrm{d}\phi ,
\end{equation}
which can be justified assuming that rings that contain a vortex will continuously resupply the vortex with dust as the latter drifts along azimuth towards the vortex, effectively converting the entire dust ring into planetesimals\footnote{This further assumes that such vortices will survive long enough for such azimuthal trapping to occur, which should be reasonably well satisfied.}.

We then compare this quantity across models, denoted as ``$M_\text{plan}$--$M_\text{plan}^\text{upper}$''. Of the $\approx\!0.38\,\Mearth$ of dust mass within the burst region in each model discussed here (and in Fig.~\ref{fig:Z}), we find that 0.12--0.19$\,\Mearth$ (32--51\%) is converted in the 1D model, 0.004--0.04$\,\Mearth$ (1--10\%) in the 2D fiducial model, $10^{-7}$--$10^{-5}\,\Mearth$ ($\approx\!0$\%) in the high-resolution model, and 0.24--0.39$\,\Mearth$ (58--96\%) in the model with $\vfrag=10\,$m/s. This further supports our conclusion that planetesimal formation during the burst phase is likely suppressed by the strong turbulent diffusion driven by the RWI, but can be efficient if dust grains grow to larger sizes. This finding is entirely consistent with the discussion around Fig.~\ref{fig:Z}, though we discuss this further in Sect.~\ref{sub:discussion-trapping}.

In summary, while the outermost pressure maximum in the post-burst region can still act as an efficient dust trap during the quiescent phase, planetesimal formation during the burst phase itself is practically quenched by the strong turbulent diffusion driven by the RWI. 
This should not interfere with planetesimal formation in the quiescent post-burst phase, however, and we expect that the total planetesimal mass of $\sim\!1\,\Mearth$ accumulating over 50\,kyr at $R\sim0.8$\,au \citepalias{ziampras-etal-2026} to remain---at least qualitatively---robust.

\section{Discussion}
\label{sec:discussion}

In this section we discuss our findings in the context of previous work. We also elaborate on their implications for the formation of planetesimals during accretion outbursts.

\subsection{More self-consistent dust growth and azimuthal trapping}
\label{sub:discussion-trapping}

In Sect.~\ref{sub:dusty-features} we showed that our models predict practically no planetesimal formation during the burst evolution itself for our highest resolution model. This was attributed to the formation of many small-scale vortices, which collectively act more akin to a source of turbulence rather than a dust trap, ultimately hindering dust accumulation on the vortices formed by the burst evolution. We nevertheless found that, were dust allowed to grow to a much larger size (with $\St\sim0.1$ in our model with $\vfrag=10$\,m/s), this issue could be sidestepped as both radial and azimuthal trapping are much more efficient.

Our results can be rationalized with simple arguments motivated by a trapping--mixing equilibrium, expressed with the ratio $\St/\alpha$ \citep[e.g.,][]{birnstiel-2024}. With typical $\St_\text{max}\sim0.003$ and $\sim\!0.1$ for our models with $\vfrag=1$ and 10\,m/s and a vortex-driven $\alpha_\text{vort}\sim10^{-2}$ for both (Fig.~\ref{fig:alpha-time}), we find that $\St/\alpha\sim0.3$ and $\sim\!10$ for the two models, respectively. This confirms that we should see marginal to weak trapping for our fiducial 2D model but noticeable trapping for the model with $\vfrag=10$\,m/s, in line with our findings. 

However, both the above estimate and our models themselves neglect the effects of vortex-driven turbulence on the dust growth process itself. Given that the \tripod{} method cannot currently capture dynamical changes in $\alpha$ due to the RWI, the accumulation of dust within vortices should be even less efficient due to the lower maximum grain size expected in the presence of strong turbulence (here $\alpha_\text{vort}\gtrsim10^{-2}$). In this context, and assuming a maximum grain size mediated by turbulent fragmentation, the largest grains will follow \citep{birnstiel-2024}
\begin{equation}
	\label{eq:St-alpha}
	\frac{\St}{\alpha}\Bigg|_\text{max} \approx \frac{1}{3\alpha^2}\frac{\vfrag^2}{\cs^2} \sim 60\,\left(\frac{10^{-4}}{\alpha}\frac{\vfrag}{1\,\text{m/s}}\right)^2\quad\text{at}~R\sim1\,\text{au}.
\end{equation}
Here, for our choice of $\alphaDZ=10^{-4}$ we should expect strong trapping in our 1D model (as $\St/\alpha\sim60\gg1$), but the turbulence generated by the vortices in our 2D model with $\alpha_\text{vort}\sim10^{-2}$ would reduce this ratio to $\St/\alpha\sim0.006\ll1$, preventing any meaningful trapping. Repeating the same calculation for our model with $\vfrag=10$\,m/s, Eq.~\eqref{eq:St-alpha} yields $\St/\alpha\sim0.6$, which lies in the same ballpark as our fiducial 2D model. This suggests that little to no planetesimal formation should occur if $\alpha_\text{vort}$ was incorporated into the dust growth prescription. As a result, we expect that the rate of planetesimal formation during an accretion outburst is negligible, even considering a rather large $\vfrag=10$\,m/s.% to very large sizes ($\St_\text{max}\sim0.1$) that can both efficiently be trapped and could even contribute to the faster dissipation of vortices via dust backreaction \citep{raettig-etal-2015,lovascio-etal-2022}.

\subsection{Comparison to 3D studies of RWI-driven vortices at the DZIE}
\label{sub:discussion-elliptical}

Our 2D models show the formation of numerous, compact vortices with aspect ratios $\chi\equiv R\Delta\phi/\Delta R$ ranging between $\approx3$ and $\approx5$ (see, e.g., Fig.~\ref{fig:vortensity-2D-vertical}). This is a direct consequence of the high vortensity contrast at the burst front forming low Rossby number vortices, which are known to be more compact \citep[see, e.g.,][]{hammer-lin-2023}. However, such compact vortices have not been observed in 3D models of the RWI near the inner rim of protoplanetary disks \citep[e.g.,][]{flock-etal-2017a,hsu-etal-2024,roberts-etal-2025,roberts-etal-2026}, where instead vortices with $\chi\sim8$ were found instead.

This discrepancy could be attributed to a number of factors. For one, our models feature a much higher resolution than those in the aforementioned works, which likely allowed us to resolve many more small-scale vortices, at the cost of assuming a 2D geometry. Secondly, the models in the above works, due to their 3D nature, are subject to the elliptical instability \citep{lesur-papaloizou-2009}, which is known to be particularly disruptive for vortices with $\chi\lesssim4$ and could thus prevent the formation of compact vortices in 3D. Nevertheless, the process of small-scale vortices forming and coalescing into a single, larger vortex in the context of the RWI acting at the DZIE has been demonstrated in 3D by \citet{roberts-etal-2025}. Finally, the compactness of vortices in our models could be caused by the very sharp vortensity gradient along the burst front, as the above studies do not actually feature a burst-like evolution but rather a static density bump at the DZIE, which would lead to a much lower (in absolute value) Rossby number and thus less compact vortices \citep[see, e.g.,][]{hammer-lin-2023}.
% Add cecil-etal-2026b after it's out

\subsection{Simplifications and caveats in our models}
\label{sub:discussion-simplifications}

Perhaps the most significant simplification in our models is the assumption of a vertically integrated disk structure. While in line with many other works in the literature on this topic \citep[e.g.,][]{lin-etal-1985,armitage-etal-2001,zhu-etal-2009,bae-etal-2013,chambers-2024}, this prevents us from making statements about the vertical structure of the disk and the development of vertical motions, with implications on the development of shadows by the directly illuminated inner rim \citep[e.g.,][]{dullemond-monnier-2010,flock-etal-2025}, the propagation of the burst front itself \citep{cecil-flock-2024}, and the growth of the elliptical instability at the core of vortices \citep{lesur-papaloizou-2009}. However, since we are primarily interested in the overall radial structure of the disk and more midplane-relevant processes such as the burst evolution, dust--gas dynamics, and planetesimal formation, our vertically integrated approach should be sufficient for our purposes. This is corroborated by the very good agreement between our 1D axisymmetric models in \citetalias{ziampras-etal-2026} and the 2D $\{R,z\}$ models of \citet{cecil-flock-2024}.

Regarding our numerical setup, we have omitted several effects that could potentially affect the details of the burst evolution, such as the feedback of the accretion flow on the stellar irradiation, more accurate treatments of the gas opacity, or the growth of the MRI on more physically motivated timescales (rather than instantly). We expect that the two former effects would not affect the burst-induced features in a meaningful way \citep[see also][]{cecil-etal-2026}, while the latter could possibly affect the vigor of the burst (Cecil et al., in prep.) and thus the strength of the RWI but without qualitatively changing our conclusions. % add Cecil et al. 2026b after it's out

Finally, the treatment of dust sublimation via a simple temperature-dependent dust-to-gas ratio \citep[following][]{cecil-flock-2024,ziampras-etal-2026} is a simplification that both assumes a single sublimation temperature for all dust species and does not capture the gradual nature of the sublimation--condensation process. Similarly to our above point on the MRI growth, we expect that a more self-consistent treatment of the sublimation of dust but also the various volatile species in the inner disk could affect the details of the burst evolution. This will be investigated in future work (Kaufmann et al., in prep.). Nevertheless, we expect that, should a burst event occur, RWI-related features will still form and affect the burst evolution in a similar way to what we have found here.

\section{Summary}
\label{sec:summary}

In this paper we have carried out very high-resolution multifluid radiation hydrodynamics simulations of the inner rim of a protoplanetary disk. We have investigated the formation and evolution of nonaxisymmetric features during an accretion outburst driven by the thermal instability, and their implications for the formation of planetesimals during the burst phase itself.

We have confirmed that the burst front is unstable to the Rossby-wave instability (RWI), which leads to the formation of a large number of small-scale vortices. These vortices coalesce into a single vortex over the course of the burst phase, but primarily act as a source of turbulence, which significantly diffuses the burst front and smooths out the serrated density profile predicted by 1D models. In doing so, the RWI-induced vortices also smooth out any variability in the accretion rate onto the star, even though the burst event itself features several reflare events.

We then quantified the strength of the vortex-driven turbulent stress, finding that it can exceed $\alpha_\text{vort}\sim10^{-2}$ over the burst region, with a radial profile that is well fit by a power law $\alpha_\text{vort}\propto R^{0.55}$, although the origin of the specific exponent remains unclear. The amplitude of $\alpha_\text{vort}$ was generally higher for lower values of $\alphaDZ$ and higher numerical resolution, but was hardly affected by dust backreaction or grain size due to the relatively low $\St$ and dust-to-gas ratios in the burst region. This high vortex-driven turbulent stress is primarily active during the burst phase itself, however, with the disk returning to a low-turbulence, quasi-axisymmetric state as the burst front recedes and vortices dissipate.

Regarding the formation of planetesimals during the burst phase, we found that the strong turbulent diffusion driven by the RWI clearly hinders dust accumulation within vortices, effectively quenching planetesimal formation over the burst cycle. This is in stark contrast to predictions of axisymmetric models \citep[e.g.,][]{cecil-flock-2024,ziampras-etal-2026}, which do not capture the RWI and its associated turbulent diffusion. However, as such vortices are expected to dissipate on rather short timescales \citep[$\sim1$\,kyr for $\alphaDZ=10^{-4}$, see][]{rometsch-etal-2021}, the resulting post-burst region should still be able to act as an efficient dust trap for inwardly drifting pebbles. This in turn suggests that planetesimals can still form in the post-burst region as predicted in \citetalias{ziampras-etal-2026}.

Aside from the detrimental effect of RWI-driven vortices on the formation of substructures and planetesimals during the burst phase, their presence does not significantly impact the overall disk profile once the burst event has passed, with the analytical predictions from \citetalias{ziampras-etal-2026} for the shape of the post-burst region being in excellent agreement with our 2D results. In particular, as vortices smear out the series of sharp density peaks seen in 1D models, the above analytical predictions are now even more relevant, with possible applications in 1D models of planet formation and migration, observable signatures of accretion outbursts, or even as an initial/inner boundary condition for global disk evolution models.

Overall, our results suggest that vortex formation is a robust feature of accretion outbursts that ultimately suppresses planetesimal formation. Vortex activity is limited to the burst phase itself, smearing out sharp radial and temporal features during its evolution but leaving the post-burst state largely unaffected. The latter enables the use of analytical predictions for the pre- and post-burst structure in 1D models of protoplanetary disks.

% \newpage
\begin{acknowledgements}
The authors would like to thank Mario Flock, Michael Cecil, and Takahiro Ueda for helpful discussions. The authors acknowledge funding from the European Union under the European Union's Horizon Europe Research and Innovation Programme 101124282 (EARLYBIRD). Views and opinions expressed are those of the authors only. TB acknowledges funding by the Deutsche Forschungsgemeinschaft (DFG, German Research Foundation) under Germany's Excellence Strategy - EXC-2094 - 390783311. Computations were performed on the HPC system Raven at the Max Planck Computing and Data Facility. This research was supported in part by grant NSF PHY-2309135 to the Kavli Institute for Theoretical Physics (KITP). All plots in this paper were made with the Python library \texttt{matplotlib} \citep{hunter-2007}.
\end{acknowledgements}

\section*{Data Availability}

Data from our numerical models are available upon reasonable request to the corresponding author.

\bibliographystyle{aa}
\bibliography{refs}

@ARTICLE{ziampras-etal-2020a,
	author = {{Ziampras}, Alexandros and {Ataiee}, Sareh and {Kley}, Wilhelm and {Dullemond}, Cornelis P. and {Baruteau}, Cl{\'e}ment},
	title = "{The impact of planet wakes on the location and shape of the water ice line in a protoplanetary disk}",
	journal = {\aap},
	year = 2020,
	month = jan,
	volume = {633},
	eid = {A29},
	pages = {A29},
	doi = {10.1051/0004-6361/201936495},
	archivePrefix = {arXiv},
	eprint = {1910.08560},
	primaryClass = {astro-ph.EP},
	adsurl = {https://ui.adsabs.harvard.edu/abs/2020A&A...633A..29Z}
}

@ARTICLE{shakura-sunyaev-1973,
       author = {{Shakura}, N.~I. and {Sunyaev}, R.~A.},
        title = "{Reprint of 1973A\&amp;A....24..337S. Black holes in binary systems. Observational appearance.}",
      journal = {\aap},
         year = "1973",
        month = "Jun",
       volume = {500},
        pages = {33-51},
       adsurl = {https://ui.adsabs.harvard.edu/\#abs/1973A&A....24..337S}
}

@ARTICLE{mignone-etal-2007,
       author = {{Mignone}, A. and {Bodo}, G. and {Massaglia}, S. and {Matsakos}, T. and
         {Tesileanu}, O. and {Zanni}, C. and {Ferrari}, A.},
        title = "{PLUTO: A Numerical Code for Computational Astrophysics}",
      journal = {The Astrophysical Journal Supplement Series},
         year = "2007",
        month = "May",
       volume = {170},
        pages = {228-242},
          doi = {10.1086/513316},
archivePrefix = {arXiv},
       eprint = {astro-ph/0701854},
 primaryClass = {astro-ph},
       adsurl = {https://ui.adsabs.harvard.edu/\#abs/2007ApJS..170..228M}
}

@ARTICLE{masset-2000,
   author = {{Masset}, F.},
    title = "{FARGO: A fast eulerian transport algorithm for differentially rotating disks}",
  journal = {\aaps},
   eprint = {astro-ph/9910390},
     year = 2000,
    month = jan,
   volume = 141,
    pages = {165-173},
      doi = {10.1051/aas:2000116},
   adsurl = {http://adsabs.harvard.edu/abs/2000A%26AS..141..165M}
}

@ARTICLE{mignone-etal-2012,
   author = {{Mignone}, A. and {Flock}, M. and {Stute}, M. and {Kolb}, S.~M. and 
	{Muscianisi}, G.},
    title = "{A conservative orbital advection scheme for simulations of magnetized shear flows with the PLUTO code}",
  journal = {\aap},
archivePrefix = "arXiv",
   eprint = {1207.2955},
 primaryClass = "astro-ph.IM",
     year = 2012,
    month = sep,
   volume = 545,
      eid = {A152},
    pages = {A152},
      doi = {10.1051/0004-6361/201219557},
   adsurl = {http://adsabs.harvard.edu/abs/2012A%26A...545A.152M}
}

@ARTICLE{bell-lin-1994,
	author = {{Bell}, K.~R. and {Lin}, D.~N.~C.},
	title = "{Using FU Orionis outbursts to constrain self-regulated protostellar disk models}",
	journal = {\apj},
	eprint = {astro-ph/9312015},
	year = 1994,
	month = jun,
	volume = 427,
	pages = {987-1004},
	doi = {10.1086/174206},
	adsurl = {http://adsabs.harvard.edu/abs/1994ApJ...427..987B}
}

@ARTICLE{balbus-papaloizou-1999,
       author = {{Balbus}, Steven A. and {Papaloizou}, John C.~B.},
        title = "{On the Dynamical Foundations of {\ensuremath{\alpha}} Disks}",
      journal = {\apj},
         year = 1999,
        month = aug,
       volume = {521},
       number = {2},
        pages = {650-658},
          doi = {10.1086/307594},
archivePrefix = {arXiv},
       eprint = {astro-ph/9903035},
 primaryClass = {astro-ph},
       adsurl = {https://ui.adsabs.harvard.edu/abs/1999ApJ...521..650B}
}

@ARTICLE{rometsch-etal-2021,
       author = {{Rometsch}, Thomas and {Ziampras}, Alexandros and {Kley}, Wilhelm and {B{\'e}thune}, William},
        title = "{Survival of planet-induced vortices in 2D disks}",
      journal = {\aap},
         year = 2021,
        month = dec,
       volume = {656},
          eid = {A130},
        pages = {A130},
          doi = {10.1051/0004-6361/202142105},
archivePrefix = {arXiv},
       eprint = {2110.00589},
 primaryClass = {astro-ph.EP},
       adsurl = {https://ui.adsabs.harvard.edu/abs/2021A&A...656A.130R}
}

@article{lovelace-1999,
	title = {Rossby {Wave} {Instability} of {Keplerian} {Accretion} {Disks}},
	volume = {513},
	issn = {0004-637X},
	url = {http://adsabs.harvard.edu/abs/1999ApJ...513..805L},
	doi = {10.1086/306900},
	urldate = {2021-02-08},
	journal = {\apj},
	author = {Lovelace, R. V. E. and Li, H. and Colgate, S. A. and Nelson, A. F.},
	month = mar,
	year = {1999},
	pages = {805--810}
}

@ARTICLE{levermore-pomraning-1981,
	author = {{Levermore}, C.~D. and {Pomraning}, G.~C.},
	title = "{A flux-limited diffusion theory}",
	journal = {\apj},
	year = 1981,
	month = aug,
	volume = {248},
	pages = {321-334},
	doi = {10.1086/159157},
	adsurl = {https://ui.adsabs.harvard.edu/abs/1981ApJ...248..321L}
}

@article{balbus-hawley-1991,
	author = {{Balbus}, Steven A. and {Hawley}, John F.},
	title = "{A Powerful Local Shear Instability in Weakly Magnetized Disks. I.
	Linear Analysis}",
	journal = {\apj},
	year = 1991,
	month = jul,
	volume = {376},
	pages = {214},
	doi = {10.1086/170270},
	adsurl = {https://ui.adsabs.harvard.edu/abs/1991ApJ...376..214B},
}

@ARTICLE{gammie-1996,
	author = {{Gammie}, Charles F.},
	title = "{Layered Accretion in T Tauri Disks}",
	journal = {\apj},
	year = 1996,
	month = jan,
	volume = {457},
	pages = {355},
	doi = {10.1086/176735},
	adsurl = {https://ui.adsabs.harvard.edu/abs/1996ApJ...457..355G}
}

@ARTICLE{hawley-etal-1995,
	author = {{Hawley}, John F. and {Gammie}, Charles F. and {Balbus}, Steven A.},
	title = "{Local Three-dimensional Magnetohydrodynamic Simulations of Accretion Disks}",
	journal = {\apj},
	year = 1995,
	month = feb,
	volume = {440},
	pages = {742},
	doi = {10.1086/175311},
	adsurl = {https://ui.adsabs.harvard.edu/abs/1995ApJ...440..742H}
}

@ARTICLE{goodman-rafikov-2001,
	author = {{Goodman}, J. and {Rafikov}, R.~R.},
	title = "{Planetary Torques as the Viscosity of Protoplanetary Disks}",
	journal = {\apj},
	year = 2001,
	month = may,
	volume = {552},
	number = {2},
	pages = {793-802},
	doi = {10.1086/320572},
	archivePrefix = {arXiv},
	eprint = {astro-ph/0010576},
	primaryClass = {astro-ph},
	adsurl = {https://ui.adsabs.harvard.edu/abs/2001ApJ...552..793G}
}

@Article{hunter-2007,
	Author    = {Hunter, J. D.},
	Title     = {Matplotlib: A 2D graphics environment},
	Journal   = {Computing In Science \& Engineering},
	Volume    = {9},
	Number    = {3},
	Pages     = {90--95},
	publisher = {IEEE COMPUTER SOC},
	year      = 2007
}

@ARTICLE{mcnally-etal-2019a,
       author = {{McNally}, Colin P. and {Nelson}, Richard P. and {Paardekooper}, Sijme-Jan and {Ben{\'\i}tez-Llambay}, Pablo},
        title = "{Migrating super-Earths in low-viscosity discs: unveiling the roles of feedback, vortices, and laminar accretion flows}",
      journal = {\mnras},
         year = 2019,
        month = mar,
       volume = {484},
       number = {1},
        pages = {728-748},
          doi = {10.1093/mnras/stz023},
archivePrefix = {arXiv},
       eprint = {1811.12841},
 primaryClass = {astro-ph.EP},
       adsurl = {https://ui.adsabs.harvard.edu/abs/2019MNRAS.484..728M}
}

@ARTICLE{bai-stone-2013b,
       author = {{Bai}, Xue-Ning and {Stone}, James M.},
        title = "{Wind-driven Accretion in Protoplanetary Disks. I. Suppression of the Magnetorotational Instability and Launching of the Magnetocentrifugal Wind}",
      journal = {\apj},
         year = 2013,
        month = may,
       volume = {769},
       number = {1},
          eid = {76},
        pages = {76},
          doi = {10.1088/0004-637X/769/1/76},
archivePrefix = {arXiv},
       eprint = {1301.0318},
 primaryClass = {astro-ph.EP},
       adsurl = {https://ui.adsabs.harvard.edu/abs/2013ApJ...769...76B}
}

@ARTICLE{birnstiel-2024,
       author = {{Birnstiel}, Tilman},
        title = "{Dust Growth and Evolution in Protoplanetary Disks}",
      journal = {\araa},
         year = 2024,
        month = sep,
       volume = {62},
       number = {1},
        pages = {157-202},
          doi = {10.1146/annurev-astro-071221-052705},
archivePrefix = {arXiv},
       eprint = {2312.13287},
 primaryClass = {astro-ph.EP},
       adsurl = {https://ui.adsabs.harvard.edu/abs/2024ARA&A..62..157B}
}

@ARTICLE{flock-etal-2017a,
       author = {{Flock}, M. and {Fromang}, S. and {Turner}, N.~J. and {Benisty}, M.},
        title = "{3D Radiation Nonideal Magnetohydrodynamical Simulations of the Inner Rim in Protoplanetary Disks}",
      journal = {\apj},
         year = 2017,
        month = feb,
       volume = {835},
       number = {2},
          eid = {230},
        pages = {230},
          doi = {10.3847/1538-4357/835/2/230},
archivePrefix = {arXiv},
       eprint = {1612.02740},
 primaryClass = {astro-ph.EP},
       adsurl = {https://ui.adsabs.harvard.edu/abs/2017ApJ...835..230F}
}

@ARTICLE{lovascio-etal-2022,
       author = {{Lovascio}, Francesco and {Paardekooper}, Sijme-Jan and {McNally}, Colin},
        title = "{Dynamics of dusty vortices - II. Stability of 2D dust-laden vortices}",
      journal = {\mnras},
         year = 2022,
        month = oct,
       volume = {516},
       number = {2},
        pages = {1635-1643},
          doi = {10.1093/mnras/stac2269},
archivePrefix = {arXiv},
       eprint = {2209.03140},
 primaryClass = {astro-ph.EP},
       adsurl = {https://ui.adsabs.harvard.edu/abs/2022MNRAS.516.1635L}
}

@software{dominik-etal-2021,
       author = {{Dominik}, Carsten and {Min}, Michiel and {Tazaki}, Ryo},
        title = "{OpTool: Command-line driven tool for creating complex dust opacities}",
 howpublished = {Astrophysics Source Code Library, record ascl:2104.010},
         year = 2021,
        month = apr,
          eid = {ascl:2104.010},
       adsurl = {https://ui.adsabs.harvard.edu/abs/2021ascl.soft04010D}
}

@ARTICLE{weber-etal-2019,
       author = {{Weber}, Philipp and {P{\'e}rez}, Sebasti{\'a}n and {Ben{\'\i}tez-Llambay}, Pablo and {Gressel}, Oliver and {Casassus}, Simon and {Krapp}, Leonardo},
        title = "{Predicting the Observational Signature of Migrating Neptune-sized Planets in Low-viscosity Disks}",
      journal = {\apj},
         year = 2019,
        month = oct,
       volume = {884},
       number = {2},
          eid = {178},
        pages = {178},
          doi = {10.3847/1538-4357/ab412f},
archivePrefix = {arXiv},
       eprint = {1909.01661},
 primaryClass = {astro-ph.EP},
       adsurl = {https://ui.adsabs.harvard.edu/abs/2019ApJ...884..178W}
}

@ARTICLE{ziampras-etal-2025a,
       author = {{Ziampras}, Alexandros and {Sudarshan}, Prakruti and {Dullemond}, Cornelis P. and {Flock}, Mario and {Berta}, Vittoria and {Nelson}, Richard P. and {Mignone}, Andrea},
        title = "{Dusty substructures induced by planets in ALMA discs: how dust growth and dynamics changes the picture}",
      journal = {\mnras},
         year = 2025,
        month = feb,
       volume = {536},
       number = {4},
        pages = {3322-3337},
          doi = {10.1093/mnras/stae2751},
archivePrefix = {arXiv},
       eprint = {2409.15420},
 primaryClass = {astro-ph.EP},
       adsurl = {https://ui.adsabs.harvard.edu/abs/2025MNRAS.536.3322Z}
}

@ARTICLE{hammer-lin-2023,
       author = {{Hammer}, Michael and {Lin}, Min-Kai},
        title = "{How to form compact and other longer-lived planet-induced vortices: VSI, planet migration, or re-triggers, but not feedback}",
      journal = {\mnras},
         year = 2023,
        month = oct,
       volume = {525},
       number = {1},
        pages = {123-149},
          doi = {10.1093/mnras/stad2264},
archivePrefix = {arXiv},
       eprint = {2304.01674},
 primaryClass = {astro-ph.EP},
       adsurl = {https://ui.adsabs.harvard.edu/abs/2023MNRAS.525..123H}
}

@ARTICLE{lesur-papaloizou-2009,
       author = {{Lesur}, G. and {Papaloizou}, J.~C.~B.},
        title = "{On the stability of elliptical vortices in accretion discs}",
      journal = {\aap},
         year = 2009,
        month = apr,
       volume = {498},
       number = {1},
        pages = {1-12},
          doi = {10.1051/0004-6361/200811577},
archivePrefix = {arXiv},
       eprint = {0903.1720},
 primaryClass = {astro-ph.EP},
       adsurl = {https://ui.adsabs.harvard.edu/abs/2009A&A...498....1L}
}

@ARTICLE{cecil-flock-2024,
       author = {{Cecil}, Michael and {Flock}, Mario},
        title = "{Variability of the inner dead zone edge in 2D radiation hydrodynamic simulations}",
      journal = {\aap},
         year = 2024,
        month = dec,
       volume = {692},
          eid = {A171},
        pages = {A171},
          doi = {10.1051/0004-6361/202451175},
archivePrefix = {arXiv},
       eprint = {2411.05444},
 primaryClass = {astro-ph.EP},
       adsurl = {https://ui.adsabs.harvard.edu/abs/2024A&A...692A.171C}
}

@ARTICLE{pfeil-etal-2024,
       author = {{Pfeil}, Thomas and {Birnstiel}, Til and {Klahr}, Hubert},
        title = "{TriPoD: Tri-Population size distributions for Dust evolution: Coagulation in vertically integrated hydrodynamic simulations of protoplanetary disks}",
      journal = {\aap},
         year = 2024,
        month = nov,
       volume = {691},
          eid = {A45},
        pages = {A45},
          doi = {10.1051/0004-6361/202449337},
archivePrefix = {arXiv},
       eprint = {2409.03816},
 primaryClass = {astro-ph.EP},
       adsurl = {https://ui.adsabs.harvard.edu/abs/2024A&A...691A..45P}
}

@INPROCEEDINGS{fischer-etal-2023,
       author = {{Fischer}, W.~J. and {Hillenbrand}, L.~A. and {Herczeg}, G.~J. and {Johnstone}, D. and {Kospal}, A. and {Dunham}, M.~M.},
        title = "{Accretion Variability as a Guide to Stellar Mass Assembly}",
    booktitle = {Protostars and Planets VII},
         year = 2023,
       editor = {{Inutsuka}, S. and {Aikawa}, Y. and {Muto}, T. and {Tomida}, K. and {Tamura}, M.},
       series = {Astronomical Society of the Pacific Conference Series},
       volume = {534},
        month = jul,
        pages = {355},
          doi = {10.48550/arXiv.2203.11257},
archivePrefix = {arXiv},
       eprint = {2203.11257},
 primaryClass = {astro-ph.SR},
       adsurl = {https://ui.adsabs.harvard.edu/abs/2023ASPC..534..355F}
}

@ARTICLE{dullemond-monnier-2010,
       author = {{Dullemond}, C.~P. and {Monnier}, J.~D.},
        title = "{The Inner Regions of Protoplanetary Disks}",
      journal = {\araa},
         year = 2010,
        month = sep,
       volume = {48},
        pages = {205-239},
          doi = {10.1146/annurev-astro-081309-130932},
archivePrefix = {arXiv},
       eprint = {1006.3485},
 primaryClass = {astro-ph.SR},
       adsurl = {https://ui.adsabs.harvard.edu/abs/2010ARA&A..48..205D}
}

@INPROCEEDINGS{lesur-etal-2023b,
       author = {{Lesur}, G. and {Flock}, M. and {Ercolano}, B. and {Lin}, M. -K. and {Yang}, C. and {Barranco}, J.~A. and {Benitez-Llambay}, P. and {Goodman}, J. and {Johansen}, A. and {Klahr}, H. and {Laibe}, G. and {Lyra}, W. and {Marcus}, P.~S. and {Nelson}, R.~P. and {Squire}, J. and {Simon}, J.~B. and {Turner}, N.~J. and {Umurhan}, O.~M. and {Youdin}, A.~N.},
        title = "{Hydro-, Magnetohydro-, and Dust-Gas Dynamics of Protoplanetary Disks}",
    booktitle = {Protostars and Planets VII},
         year = 2023,
       editor = {{Inutsuka}, S. and {Aikawa}, Y. and {Muto}, T. and {Tomida}, K. and {Tamura}, M.},
       series = {Astronomical Society of the Pacific Conference Series},
       volume = {534},
        month = jul,
        pages = {465},
          doi = {10.48550/arXiv.2203.09821},
archivePrefix = {arXiv},
       eprint = {2203.09821},
 primaryClass = {astro-ph.EP},
       adsurl = {https://ui.adsabs.harvard.edu/abs/2023ASPC..534..465L}
}

@ARTICLE{armitage-etal-2001,
       author = {{Armitage}, Philip J. and {Livio}, Mario and {Pringle}, J.~E.},
        title = "{Episodic accretion in magnetically layered protoplanetary discs}",
      journal = {\mnras},
         year = 2001,
        month = jun,
       volume = {324},
       number = {3},
        pages = {705-711},
          doi = {10.1046/j.1365-8711.2001.04356.x},
archivePrefix = {arXiv},
       eprint = {astro-ph/0101253},
 primaryClass = {astro-ph},
       adsurl = {https://ui.adsabs.harvard.edu/abs/2001MNRAS.324..705A}
}

@ARTICLE{zhu-etal-2009,
       author = {{Zhu}, Zhaohuan and {Hartmann}, Lee and {Gammie}, Charles},
        title = "{Nonsteady Accretion in Protostars}",
      journal = {\apj},
         year = 2009,
        month = apr,
       volume = {694},
       number = {2},
        pages = {1045-1055},
          doi = {10.1088/0004-637X/694/2/1045},
archivePrefix = {arXiv},
       eprint = {0811.1762},
 primaryClass = {astro-ph},
       adsurl = {https://ui.adsabs.harvard.edu/abs/2009ApJ...694.1045Z}
}

@ARTICLE{chambers-2024,
       author = {{Chambers}, John},
        title = "{Large Fluctuations within 1 au in Protoplanetary Disks}",
      journal = {\apj},
         year = 2024,
        month = may,
       volume = {966},
       number = {1},
          eid = {40},
        pages = {40},
          doi = {10.3847/1538-4357/ad3731},
archivePrefix = {arXiv},
       eprint = {2403.17126},
 primaryClass = {astro-ph.EP},
       adsurl = {https://ui.adsabs.harvard.edu/abs/2024ApJ...966...40C}
}

@ARTICLE{wunsch-etal-2006,
       author = {{W{\"u}nsch}, R. and {Gawryszczak}, A. and {Klahr}, H. and {R{\'o}{\.z}yczka}, M.},
        title = "{Two-dimensional models of layered protoplanetary discs - II. The effect of a residual viscosity in the dead zone}",
      journal = {\mnras},
         year = 2006,
        month = apr,
       volume = {367},
       number = {2},
        pages = {773-780},
          doi = {10.1111/j.1365-2966.2005.09986.x},
archivePrefix = {arXiv},
       eprint = {astro-ph/0512402},
 primaryClass = {astro-ph},
       adsurl = {https://ui.adsabs.harvard.edu/abs/2006MNRAS.367..773W}
}

@ARTICLE{hartmann-kenyon-1996,
       author = {{Hartmann}, Lee and {Kenyon}, Scott J.},
        title = "{The FU Orionis Phenomenon}",
      journal = {\araa},
         year = 1996,
        month = jan,
       volume = {34},
        pages = {207-240},
          doi = {10.1146/annurev.astro.34.1.207},
       adsurl = {https://ui.adsabs.harvard.edu/abs/1996ARA&A..34..207H}
}

@INPROCEEDINGS{audard-etal-2014,
       author = {{Audard}, M. and {{\'A}brah{\'a}m}, P. and {Dunham}, M.~M. and {Green}, J.~D. and {Grosso}, N. and {Hamaguchi}, K. and {Kastner}, J.~H. and {K{\'o}sp{\'a}l}, {\'A}. and {Lodato}, G. and {Romanova}, M.~M. and {Skinner}, S.~L. and {Vorobyov}, E.~I. and {Zhu}, Z.},
        title = "{Episodic Accretion in Young Stars}",
    booktitle = {Protostars and Planets VI},
         year = 2014,
       editor = {{Beuther}, Henrik and {Klessen}, Ralf S. and {Dullemond}, Cornelis P. and {Henning}, Thomas},
        month = jan,
        pages = {387-410},
          doi = {10.2458/azu_uapress_9780816531240-ch017},
archivePrefix = {arXiv},
       eprint = {1401.3368},
 primaryClass = {astro-ph.SR},
       adsurl = {https://ui.adsabs.harvard.edu/abs/2014prpl.conf..387A}
}

@ARTICLE{kuznetsova-etal-2022,
       author = {{Kuznetsova}, Aleksandra and {Bae}, Jaehan and {Hartmann}, Lee and {Mac Low}, Mordecai-Mark},
        title = "{Anisotropic Infall and Substructure Formation in Embedded Disks}",
      journal = {\apj},
         year = 2022,
        month = mar,
       volume = {928},
       number = {1},
          eid = {92},
        pages = {92},
          doi = {10.3847/1538-4357/ac54a8},
archivePrefix = {arXiv},
       eprint = {2202.05301},
 primaryClass = {astro-ph.EP},
       adsurl = {https://ui.adsabs.harvard.edu/abs/2022ApJ...928...92K}
}

@ARTICLE{raettig-etal-2015,
       author = {{Raettig}, Natalie and {Klahr}, Hubert and {Lyra}, Wladimir},
        title = "{Particle Trapping and Streaming Instability in Vortices in Protoplanetary Disks}",
      journal = {\apj},
         year = 2015,
        month = may,
       volume = {804},
       number = {1},
          eid = {35},
        pages = {35},
          doi = {10.1088/0004-637X/804/1/35},
archivePrefix = {arXiv},
       eprint = {1501.05364},
 primaryClass = {astro-ph.EP},
       adsurl = {https://ui.adsabs.harvard.edu/abs/2015ApJ...804...35R}
}

@ARTICLE{flock-etal-2025,
       author = {{Flock}, Mario and {Chrenko}, Ond{\v{r}}ej and {Ueda}, Takahiro and {Benisty}, Myriam and {Varga}, Jozsef and {van Boekel}, Roy},
        title = "{Effect of multi-dust species on the inner rim of magnetized protoplanetary disks}",
      journal = {\aap},
         year = 2025,
        month = sep,
       volume = {701},
          eid = {A259},
        pages = {A259},
          doi = {10.1051/0004-6361/202453124},
archivePrefix = {arXiv},
       eprint = {2508.04254},
 primaryClass = {astro-ph.EP},
       adsurl = {https://ui.adsabs.harvard.edu/abs/2025A&A...701A.259F}
}

@ARTICLE{bae-etal-2013,
       author = {{Bae}, Jaehan and {Hartmann}, Lee and {Zhu}, Zhaohuan and {Gammie}, Charles},
        title = "{Variable Accretion Outbursts in Protostellar Evolution}",
      journal = {\apj},
         year = 2013,
        month = feb,
       volume = {764},
       number = {2},
          eid = {141},
        pages = {141},
          doi = {10.1088/0004-637X/764/2/141},
archivePrefix = {arXiv},
       eprint = {1212.6454},
 primaryClass = {astro-ph.SR},
       adsurl = {https://ui.adsabs.harvard.edu/abs/2013ApJ...764..141B}
}

@ARTICLE{bae-etal-2014,
       author = {{Bae}, Jaehan and {Hartmann}, Lee and {Zhu}, Zhaohuan and {Nelson}, Richard P.},
        title = "{Accretion Outbursts in Self-gravitating Protoplanetary Disks}",
      journal = {\apj},
         year = 2014,
        month = nov,
       volume = {795},
       number = {1},
          eid = {61},
        pages = {61},
          doi = {10.1088/0004-637X/795/1/61},
archivePrefix = {arXiv},
       eprint = {1409.3891},
 primaryClass = {astro-ph.SR},
       adsurl = {https://ui.adsabs.harvard.edu/abs/2014ApJ...795...61B}
}

@ARTICLE{Lim2025,
       author = {{Lim}, Jeonghoon and {Simon}, Jacob B. and {Li}, Rixin and {Brouillette}, Olivia and {Rea}, David G. and {Lyra}, Wladimir},
        title = "{The Streaming Instability in 3D: Conditions for Strong Clumping}",
      journal = {arXiv e-prints},
         year = 2025,
        month = sep,
          eid = {arXiv:2509.18270},
        pages = {arXiv:2509.18270},
          doi = {10.48550/arXiv.2509.18270},
archivePrefix = {arXiv},
       eprint = {2509.18270},
 primaryClass = {astro-ph.EP},
       adsurl = {https://ui.adsabs.harvard.edu/abs/2025arXiv250918270L}
}

@ARTICLE{cecil-etal-2026,
       author = {{Cecil}, Michael and {Flock}, Mario and {Malygin}, Mykola G. and {Kuiper}, Rolf and {Sudarshan}, Prakruti and {Ziampras}, Alexandros and {Elbakyan}, Vardan G.},
        title = "{The role of detailed gas and dust opacities in shaping the evolution of the inner disc edge subject to episodic accretion}",
      journal = {arXiv e-prints},
         year = 2026,
        month = feb,
          eid = {arXiv:2602.10674},
        pages = {arXiv:2602.10674},
          doi = {10.48550/arXiv.2602.10674},
archivePrefix = {arXiv},
       eprint = {2602.10674},
 primaryClass = {astro-ph.EP},
       adsurl = {https://ui.adsabs.harvard.edu/abs/2026arXiv260210674C}
}

@ARTICLE{ziampras-etal-2026,
       author = {{Ziampras}, Alexandros and {Birnstiel}, Tilman and {Kaufmann}, Nicolas and {Cecil}, Michael and {Pfeil}, Thomas},
        title = "{Planet formation at the inner edge of the dead zone -- I: the interplay between accretion outbursts and dust growth}",
      journal = {arXiv e-prints},
         year = 2026,
        month = feb,
          eid = {arXiv:2602.20283},
        pages = {arXiv:2602.20283},
          doi = {10.48550/arXiv.2602.20283},
archivePrefix = {arXiv},
       eprint = {2602.20283},
 primaryClass = {astro-ph.EP},
       adsurl = {https://ui.adsabs.harvard.edu/abs/2026arXiv260220283Z}
}

@ARTICLE{cordwell-etal-2025,
       author = {{Cordwell}, Amelia J. and {Ziampras}, Alexandros and {Brown}, Joshua J. and {Rafikov}, Roman R.},
        title = "{How two-dimensional are planet─disc interactions? I. Locally isothermal discs}",
      journal = {\mnras},
         year = 2025,
        month = nov,
       volume = {543},
       number = {4},
        pages = {4198-4217},
          doi = {10.1093/mnras/staf1674},
archivePrefix = {arXiv},
       eprint = {2509.04282},
 primaryClass = {astro-ph.EP},
       adsurl = {https://ui.adsabs.harvard.edu/abs/2025MNRAS.543.4198C}
}

@ARTICLE{birnstiel-etal-2013,
       author = {{Birnstiel}, T. and {Dullemond}, C.~P. and {Pinilla}, P.},
        title = "{Lopsided dust rings in transition disks}",
      journal = {\aap},
         year = 2013,
        month = feb,
       volume = {550},
          eid = {L8},
        pages = {L8},
          doi = {10.1051/0004-6361/201220847},
archivePrefix = {arXiv},
       eprint = {1301.1976},
 primaryClass = {astro-ph.EP},
       adsurl = {https://ui.adsabs.harvard.edu/abs/2013A&A...550L...8B}
}

@ARTICLE{lyra-lin-2013,
       author = {{Lyra}, Wladimir and {Lin}, Min-Kai},
        title = "{Steady State Dust Distributions in Disk Vortices: Observational Predictions and Applications to Transitional Disks}",
      journal = {\apj},
         year = 2013,
        month = sep,
       volume = {775},
       number = {1},
          eid = {17},
        pages = {17},
          doi = {10.1088/0004-637X/775/1/17},
archivePrefix = {arXiv},
       eprint = {1307.3770},
 primaryClass = {astro-ph.EP},
       adsurl = {https://ui.adsabs.harvard.edu/abs/2013ApJ...775...17L}
}

@ARTICLE{lin-etal-1985,
       author = {{Lin}, D.~N.~C. and {Papaloizou}, J. and {Faulkner}, J.},
        title = "{On the evolution of accretion disc flow in cataclysmic variables - III. Outburst properties of constant and uniform -alpha model discs.}",
      journal = {\mnras},
         year = 1985,
        month = jan,
       volume = {212},
        pages = {105-149},
          doi = {10.1093/mnras/212.1.105},
       adsurl = {https://ui.adsabs.harvard.edu/abs/1985MNRAS.212..105L}
}

@ARTICLE{kley-lin-1999,
       author = {{Kley}, W. and {Lin}, D.~N.~C.},
        title = "{Evolution of FU Orionis Outbursts in Protostellar Disks}",
      journal = {\apj},
         year = 1999,
        month = jun,
       volume = {518},
       number = {2},
        pages = {833-847},
          doi = {10.1086/307296},
       adsurl = {https://ui.adsabs.harvard.edu/abs/1999ApJ...518..833K}
}

@software{ziampras-birnstiel-2026,
       author = {{Ziampras}, Alexandros and {Birnstiel}, Tilman},
        title = "{growpacity: A computationally efficient dust opacity model suitable for coagulation models}",
 howpublished = {Astrophysics Source Code Library, record ascl:2603.020},
         year = 2026,
        month = mar,
          eid = {ascl:2603.020},
archivePrefix = {ascl},
       eprint = {2603.020},
       adsurl = {https://ui.adsabs.harvard.edu/abs/2026ascl.soft03020Z}
}

@ARTICLE{roberts-etal-2025,
       author = {{Roberts}, Matthew J.~O. and {Latter}, Henrik N. and {Lesur}, Geoffroy},
        title = "{Global magnetohydrodynamic simulations of the inner regions of protoplanetary discs. I. Zero-net flux regime}",
      journal = {\mnras},
         year = 2025,
        month = nov,
       volume = {544},
       number = {1},
        pages = {1284-1303},
          doi = {10.1093/mnras/staf1672},
archivePrefix = {arXiv},
       eprint = {2506.16945},
 primaryClass = {astro-ph.EP},
       adsurl = {https://ui.adsabs.harvard.edu/abs/2025MNRAS.544.1284R}
}

@ARTICLE{roberts-etal-2026,
       author = {{Roberts}, Matthew J.~O. and {Latter}, Henrik N. and {Lesur}, Geoffroy},
        title = "{Global magnetohydrodynamic simulations of the inner regions of protoplanetary discs. II. Vertical-net-flux regime}",
      journal = {\mnras},
         year = 2026,
        month = mar,
          doi = {10.1093/mnras/stag609},
archivePrefix = {arXiv},
       eprint = {2602.11818},
 primaryClass = {astro-ph.EP},
       adsurl = {https://ui.adsabs.harvard.edu/abs/2026MNRAS.tmp..574R}
}

@ARTICLE{hsu-etal-2024,
       author = {{Hsu}, Chun-Yen and {Li}, Zhi-Yun and {Tu}, Yisheng and {Hu}, Xiao and {Lin}, Min-Kai},
        title = "{Rossby wave instability and substructure formation in 3D non-ideal MHD wind-launching discs}",
      journal = {\mnras},
         year = 2024,
        month = sep,
       volume = {533},
       number = {3},
        pages = {2980-2996},
          doi = {10.1093/mnras/stae1986},
archivePrefix = {arXiv},
       eprint = {2407.08032},
 primaryClass = {astro-ph.EP},
       adsurl = {https://ui.adsabs.harvard.edu/abs/2024MNRAS.533.2980H}
}

@ARTICLE{Barge1995,
       author = {{Barge}, P. and {Sommeria}, J.},
        title = "{Did planet formation begin inside persistent gaseous vortices?}",
      journal = {\aap},
         year = 1995,
        month = mar,
       volume = {295},
        pages = {L1-L4},
          doi = {10.48550/arXiv.astro-ph/9501050},
archivePrefix = {arXiv},
       eprint = {astro-ph/9501050},
 primaryClass = {astro-ph},
       adsurl = {https://ui.adsabs.harvard.edu/abs/1995A&A...295L...1B}
}

@ARTICLE{drazkowska-etal-2016,
       author = {{Dr{\k{a}}{\.z}kowska}, J. and {Alibert}, Y. and {Moore}, B.},
        title = "{Close-in planetesimal formation by pile-up of drifting pebbles}",
      journal = {\aap},
         year = 2016,
        month = oct,
       volume = {594},
          eid = {A105},
        pages = {A105},
          doi = {10.1051/0004-6361/201628983},
archivePrefix = {arXiv},
       eprint = {1607.05734},
 primaryClass = {astro-ph.EP},
       adsurl = {https://ui.adsabs.harvard.edu/abs/2016A&A...594A.105D}
}

@ARTICLE{klahr-henning-1997,
       author = {{Klahr}, H. Hubertus and {Henning}, Thomas},
        title = "{Particle-Trapping Eddies in Protoplanetary Accretion Disks}",
      journal = {\icarus},
         year = 1997,
        month = jul,
       volume = {128},
       number = {1},
        pages = {213-229},
          doi = {10.1006/icar.1997.5720},
       adsurl = {https://ui.adsabs.harvard.edu/abs/1997Icar..128..213K}
}

@ARTICLE{guenther-etal-2018,
       author = {{G{\"u}nther}, Hans Moritz and {Birnstiel}, T. and {Huenemoerder}, D.~P. and {Principe}, D.~A. and {Schneider}, P.~C. and {Wolk}, S.~J. and {Dubois}, Franky and {Logie}, Ludwig and {Rau}, Steve and {Vanaverbeke}, Sigfried},
        title = "{Optical Dimming of RW Aur Associated with an Iron-rich Corona and Exceptionally High Absorbing Column Density}",
      journal = {\aj},
         year = 2018,
        month = aug,
       volume = {156},
       number = {2},
          eid = {56},
        pages = {56},
          doi = {10.3847/1538-3881/aac9bd},
archivePrefix = {arXiv},
       eprint = {1807.06995},
 primaryClass = {astro-ph.SR},
       adsurl = {https://ui.adsabs.harvard.edu/abs/2018AJ....156...56G}
}

% \clearpage
% \newpage

\appendix

\section{Comparison to a model without FLD}
\label{apdx:noFLD}

To assess whether radiative diffusion plays a significant role in the formation and lifetime of vortices in our models, we ran a variant of our fiducial model where we omitted in-plane radiation transport (FLD), while still including vertical cooling. We find that the overall behavior of the burst evolution, the growth of the RWI, and the resulting vortex-driven turbulent stress are very similar between the two models. The former observation is backed by our findings in \citetalias{ziampras-etal-2026}, where we had found that in-plane radiation transport was not essential for the overall radial dynamics.

However, the model without FLD shows slightly higher values of $\alpha_\text{vort}$ overall (by $\sim\!5\%$), likely due to radiative diffusion smoothing out the vortensity contrast both along the burst front and across vortices\footnote{The role of FLD in shortening vortex lifetimes has been confirmed in preliminary, currently unpublished work by A.~Ziampras.}. This can also be inferred from Fig.~\ref{fig:alpha-Rt-noFLD}, where we show the radius--time heatmap of $\bar{\alpha}_\text{turb}$ for the two models similar to Fig.~\ref{fig:alpha-Rt-gallery}. Here, the model without FLD shows a higher overall amplitude of RWI-driven turbulence (indicated by brighter colors) and slightly longer-lived vortices, but with otherwise very similar behavior.

All in all, we find that the inclusion of FLD results in slightly weaker RWI turbulence and shorter-lived vortices, but does not substantially change the overall burst evolution besides slightly delaying the onset of the first reflare in our models. This suggests that in-plane cooling is not essential for the overall burst evolution \citepalias[in line with][]{ziampras-etal-2026}, but can have a non-negligible effect on vortex-related dynamics.

\begin{figure}
	\includegraphics[width=\columnwidth]{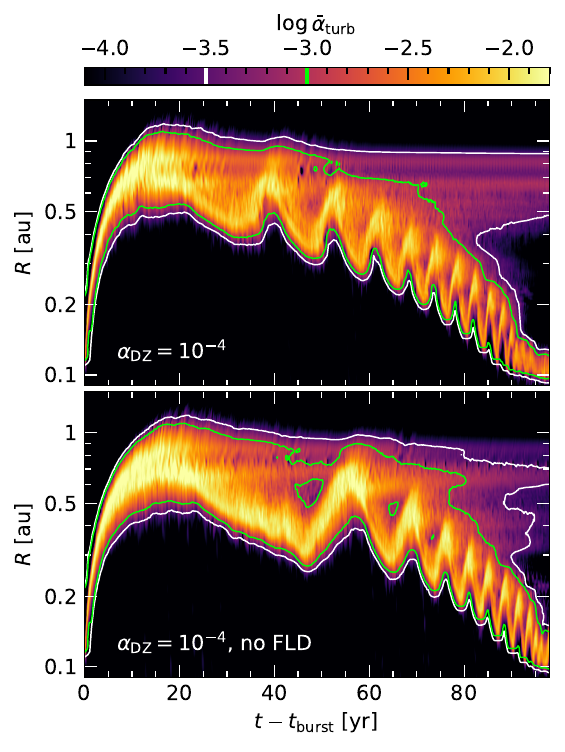}
	\caption{Radius--time heatmap of the azimuthally averaged turbulent viscosity $\bar{\alpha}_\text{turb}(R,t)$ for our fiducial model and one without FLD, similar to Fig.~\ref{fig:alpha-Rt-gallery}. The model without FLD shows a higher overall amplitude of RWI-driven turbulence and slightly longer-lived vortices, but with otherwise very similar behavior.}
	\label{fig:alpha-Rt-noFLD}
\end{figure}

\section{Comparison to a high-resolution model} \label{apdx:highres}

\begin{figure*}[t]
	\includegraphics[width=\textwidth]{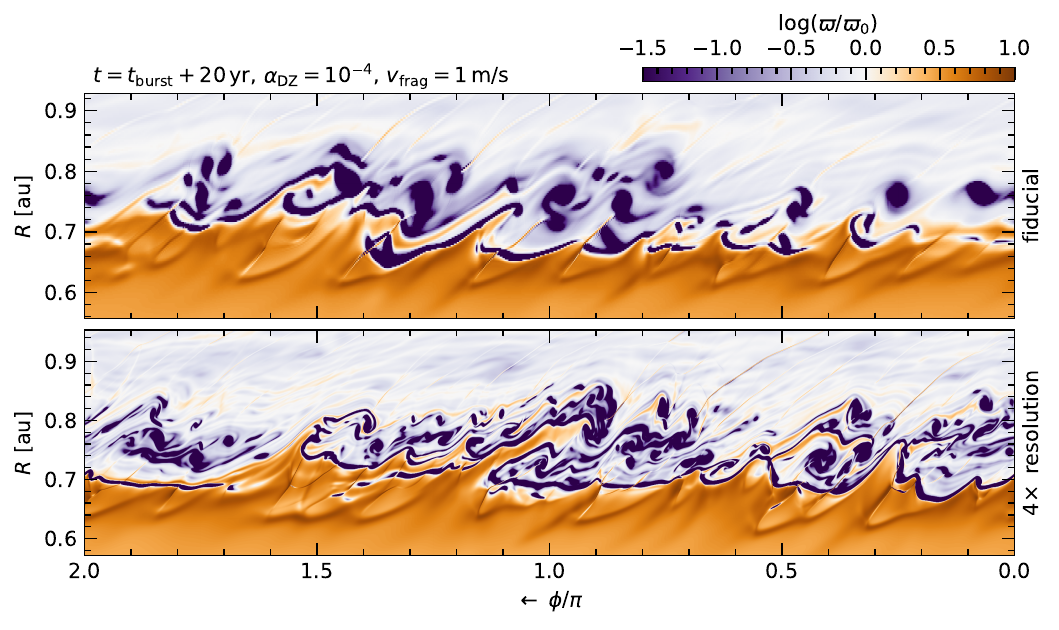}
	\caption{Heatmap of the perturbed vortensity during the time when the burst has reached its outermost point, for our fiducial (top) and high-resolution models (bottom). While the two capture the burst front in a very similar manner, the latter shows much richer azimuthal features in the form of many small-scale vortices and sharper gradients.}
	\label{fig:vortensity-2D}
\end{figure*}

\begin{figure}
	\includegraphics[width=\columnwidth]{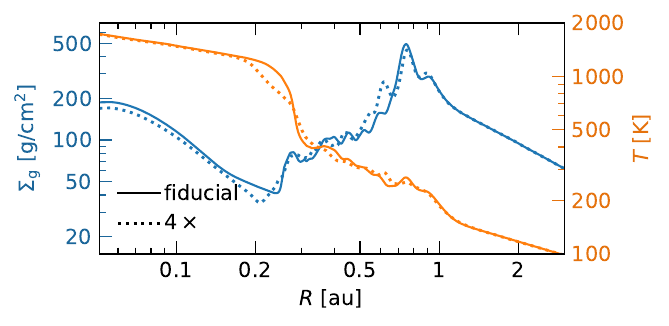}
	\caption{Comparison of the gas surface density (blue) and temperature (orange) between our fiducial (solid) and high-resolution model (dotted) during the evolution of the burst. The two models agree exceptionally well.}
	\label{fig:resolution-1D}
\end{figure}

Carrying out a comparison model at such high numerical resolution was motivated by the need to both check for numerical convergence of our fiducial model and to investigate the growth of the RWI and the resulting vortices in more detail. Indeed, we find a significantly higher number of smaller vortices in the high-resolution model, which are not fully resolved in the fiducial model and contribute to a higher overall amplitude of $\alpha_\text{vort}$ (see Sect.~\ref{sub:vortex-stress} and Fig.~\ref{fig:alpha-time}). A comparison of the vortensity field between the fiducial and high-resolution models at $t=20$\,yr is shown in Fig.~\ref{fig:vortensity-2D}, at a time when the burst front has reached its outermost radius. Interestingly, while many more vortices are resolved in the high-resolution model, the overall structure of the burst front is very similar between the two models in terms of the amplitude of vortensity perturbations, the radial width of the burst front, and its wavy azimuthal structure. 

The similarities extend to the timescale of the burst cycle, the radial structure of the disk, and the presence of multiple reflares. We showcase this in Fig.~\ref{fig:resolution-1D}, where we compare the gas density and temperature profiles at an arbitrary time well within the burst evolution. This confirms that our fiducial model is already converged acceptably well, and that our main results regarding the burst evolution and the growth of the RWI are robust against resolution. We therefore use our fiducial model as our main reference model in this paper.

\end{document}